\documentclass[longauthor]{aastex62a}
\usepackage{epsfig}
\usepackage{amsmath}
\usepackage{mathtools}

\usepackage{titlesec}
\titleformat{\section}[block]{\large\bfseries\filcenter}{\thesection.}{0.3em}{}

\hypersetup{linkcolor=black,citecolor=black,filecolor=black,urlcolor=black}

\def\cm{{\rm cm}}
\def\oday{{\rm day}}
\def\eff{{\rm eff}}

\def\es{{\rm es}}

\def\expp{{\rm exp}}
\def\g{{\rm g}}
\def\K{{\rm K}}
\def\lex{{\rm lex}}
\def\nm{{\rm nm}}
\def\s{{\rm s}}
\def\Th{{\rm Th}}

\newcommand{\llangle}{\left\langle}
\newcommand{\rrangle}{\right\rangle}
\newcommand{\arctanh}{\rm arctanh}
\newcommand{\klcs}{\rm KLCS}

\begin{document}

\title{\bf\Large Line Expansion Opacity in Relativistically Expanding Media}


\author{\normalsize Li-Xin Li}
\affiliation{\vspace{0pt}\\
  {\small Kavli Institute for Astronomy and Astrophysics, Peking University, Beijing 100871, P. R. China\\
{\rm Email: lxl@pku.edu.cn}}}

\begin{abstract}
\noindent Spectral lines of heavy atomic elements in the ejecta of supernovae and neutron star mergers can have important contribution to the opacity of the ejecta matter even when the abundance of the elements is very small. Under favorable conditions, the line expansion opacity arising from spectral lines and the expansion of the medium can be orders of magnitude larger than the opacity of electron scattering. In this paper we derive the formulae for evaluating the line expansion opacity and its Rosseland mean in an expanding medium in the framework of special relativity, which can be considered as a generalization of the previous work in the Newtonian approximation. Then we compare the derived relativistic formulae to the Newtonian ones to explore the relativistic effect on the opacity, and test the new formulae with the spectral lines of some heavy atomic elements. We also derive some approximation formulae for the Rosseland mean of the line expansion opacity that are easy to use in numerical works while still maintaining a high enough accuracy relative to exact solutions. The formulae derived in this paper are expected to have important applications in radiative problems related to relativistic astrophysical phenomena such as neutron star mergers, supernovae, and gamma-ray bursts where relativistic or subrelativistic expansions are involved. 
  
\vspace{0.3cm}
\end{abstract}


\maketitle

\vspace{0.3cm}

\section{Introduction}
\label{intro}

The opacity arising from atomic bound-bound transitions of heavy elements plays a very important role in the radiative transfer and thermodynamic equilibrium in a rapidly expanding medium, even when the medium contains only a small abundance of heavy elements. The differential Doppler effect in the expanding matter causes the intrinsically discrete and narrow spectral lines arising from atomic bound-bound transitions to be broadened and merge, resulting in an effective and quasi-continuous absorption opacity. This effective opacity, which is often named the line expansion opacity, can in many situations be larger than the ordinary continuous opacity such as the opacity arising from electron scattering. Therefore, a precise and accurate evaluation of the line expansion opacity in an expanding medium is crucial for predicting or interpreting the spectra and light curves of relevant astronomical phenomena, such as supernovae, gamma-ray bursts, and neutron star mergers.

The effect of spectral lines on the opacity in an expanding medium was systematically studied in a landmark paper by Karp et al. (1977; hereafter KLCS), where the line expansion opacity was derived from the mean free path defined by averaging the distance traveled by a continuously redshifted photon that passes by the spectral lines along its path. The research has been a theoretical basis for calculating the effective opacity arising from atomic bound-bound transitions in the expanding shell of supernovae for decades \citep{hoe93,hoe98,ber11,rab11,leu15}, although a simplified and revised version has also been widely applied for supernovae \citep{eas93,pin00,kas06a,kas06b,wyg19} and neutron star mergers recently \citep{bar13,kas13,tan13,fon17,wol18,gai19,tan19}.

A major shortcoming in the formalism of \citet{kar77} is that the formulae were derived in the framework of Newtonian mechanics, which means that the derived line expansion opacity applies only to a system with an expansion velocity much smaller than the speed of light. In the treatment of \citet{kar77}, the Newtonian approximation is directly related to the assumption that all the involved spectral lines have frequencies very close to the frequency of the test photon. In reality, relativistically expanding systems are common in astronomy, including gamma-ray bursts, neutron star mergers, and broad-line supernovae. For gamma-ray bursts, the ejecta generating the gamma-ray emission can have an expansion velocity so close to the speed of light that the corresponding Lorentz factor is $\ga 100$ \citep[][and references therein]{pir04,zha04,mes06}. For neutron star mergers (neutron star-neutron star merger and black hole-neutron star merger), the expansion speed of the the associated ejecta can be $\ga 0.3$ times the speed of light \citep[][and references therein]{lat74,lat76,li98,bau13,wax18}. For broad-line supernovae, the expansion velocity at the photosphere deduced from the observed spectral lines can be $\ga 0.1$ times the speed of light \citep{iwa98,pat01,maz03,mod06,pia06,cor16}. To calculate the line expansion opacity in these relativistically or subrelativistically expanding systems, the formalism developed in \citet{kar77} must be generalized to include the effect of special relativity.

After the discovery of the first ever event of a neutron star merger in both gravitational and electromagnetic wave bands, i.e., GW170817/GRB170817A and the optical counterpart SSS17a/AT2017gfo \citep{abb17a,abb17b,cou17,gol17,sav17,sie17,val17}, the need for a relativistic expression of the line expansion opacity becomes more urgent. It is almost evident that at early time the merger ejecta producing the SSS17a/AT2017gfo kilonova emission has an expansion velocity $\ga 0.3$ times the speed of light \citep{wax18}, and it has been claimed that the presence of even a small fraction of lanthanides in the merger ejecta can boost the opacity up to a value larger than that of electron scattering by orders of magnitude due to the bound-bound transition of the $f$-shell electrons of lanthanides \citep{kas13,tan13,tan19}. An accurate evaluation of the line expansion opacity in a relativistic ejecta is very important for determining where the optical emission generated by a neutron star merger peaks (in the ultraviolet band or in the infrared band, for instance), how bright the emission is, and how the spectrum evolves with time.

In this paper we derive the line expansion opacity in an expanding medium in the framework of special relativity and evaluate its Rosseland mean. For simplicity, we assume a uniformly expanding sphere with a uniform mass density. That is, the expansion velocity is proportional to the distance to the center of the sphere, and the mass density is constant in space in a frame comoving with the expansion. The expansion velocity at the surface of the sphere can take any value between zero and the speed of light. Following the procedure in \citet{kar77} with some necessary modifications, we derive an expression for evaluating the line expansion opacity arising from a sample of atomic spectral lines. In the derivation of the formula we relax the assumption of diffusion approximation so that the obtained result is more general than that derived in \citet{kar77}. In addition, in the definition of the mean free path from which the line expansion opacity is derived, we include the contribution from a universal expansion opacity in order to maintain the convergence of the integral. We then evaluate the Rosseland mean of the derived line expansion opacity, and test the derived formulae with the data of some spectral lines of iron and neodymium.

We will also compare the opacity result evaluated with our generalized formulae to that evaluated with the formulae of \citet{kar77}. Based on the comparison we will argue for the advantages of our formalism over others, and then discuss generalization of our formulae to a more general case---a sphere with a nonuniform mass density.

\section{Line Opacity and the Optical Depth}
\label{l_opc}

Let us consider a uniformly expanding sphere with an expansion speed $V=\beta_R c$ at its surface, where $c$ is the speed of light and $0<\beta_R<1$. The sphere has a uniform mass density $\rho$ measured in a frame comoving with the expansion of the sphere.  With the gravity of the sphere being ignored, the spacetime metric inside the sphere is given by that of the Milne universe \citep{mil35,bir82,li07,li13}. Roughly speaking, the spacetime inside the sphere is a universe with a negative spatial curvature, a vanishing Riemann curvature, and a boundary at the sphere surface. The time coordinate in the comoving frame, $\eta$, is related to the time in a rest inertial frame, $t$, by the relation $t=\eta\cosh\xi$, where $\xi$ is a dimensionless radial coordinate in the comoving frame, which is related to the radius $r$ in the inertial frame by the relation $r=c\eta\sinh\xi$. The coordinate $\xi$ is related to the {\it in situ} expansion velocity of the sphere, $v=\beta c$ where $0\le\beta\le\beta_R$, by $\xi=\arctanh\beta$. Hence, we have the Lorentz factor $\gamma\equiv(1-\beta^2)^{-1/2}=\cosh\xi$, and $t=\gamma\eta$.

Unless otherwise stated, all quantities used in this paper are defined in a frame comoving with the expansion of the sphere (the comoving frame), or equivalently, in a local rest frame of the expanding sphere.

Consider two levels of an atom labeled by numbers 1 and 2, respectively. The level 1 has an energy $E_1$ and a statistical weight $g_1$. The level 2 has an energy $E_2>E_1$ and a statistical weight $g_2$. The atom can transit from level 1 to level 2 by absorption of a photon of energy $h\nu_l=E_2-E_1$, a process described by an absorption oscillator strength $f_{12}>0$. Here $h$ is the Planck constant, and $\nu_l$ is the line frequency. Similarly, the atom can transit from level 2 to level 1 by emission of a photon of energy $h\nu_l$, a process described by an emission oscillator strength $f_{21}<0$. By the detailed balance equation, the absorption oscillator strength and the emission oscillator strength are related by $g_1f_{12}=-g_2f_{21}$. According to quantum mechanics, the absorption coefficient arising from a line absorption is given by \citep[after correction for the stimulated emission,][]{kar77,ryb04} 
\begin{eqnarray}
  \alpha_\nu = \frac{\pi e^2}{m_ec}f_{12} n_1\left(1-\frac{g_1n_2}{g_2n_1}\right)\phi(\nu) \;, 
\end{eqnarray}
where $e$ is the electron charge, $m_e$ the electron mass, $n_1$ the number density of atoms in level 1, and $n_2$ the number density of atoms in level 2. The function $\phi(\nu)$ is a line profile function sharply peaked at the frequency $\nu=\nu_l$ and normalized by the condition $\int\phi(\nu)d\nu=1$. 

If the system is in locally thermodynamic equilibrium, we have $n_2/n_1=(g_2/g_1)e^{-h\nu_l/kT}$, where $T$ is the temperature of the matter and $k$ is the Boltzmann constant. Then we get
\begin{eqnarray}
  \alpha_\nu = \frac{\pi e^2}{m_ec}f_{12} n_1\left(1-e^{-h\nu_l/kT}\right)\phi(\nu) \;. \label{alp_nu}
\end{eqnarray}

Consider a photon of frequency $\nu_0$ at time $\eta_0$ inside the sphere. As time goes on with the expansion of the sphere, the frequency of the photon measured in a local rest frame gets redshifted according to the relation $\nu\propto\eta^{-1}$. If $\nu_0<\nu_l$, the frequency of the photon gets farther away from the line frequency and will never match the line frequency. Hence, when $\nu_0<\nu_l$, we have the optical depth of the line to the photon: $\tau_l=0$. On the other hand, if $\nu_0>\nu_l$, the frequency of the photon will match the frequency of the line at time $\eta_l=\eta_0\nu_0/\nu_l$. Then we can calculate the optical depth of the line to the photon
\begin{eqnarray}
  \tau_l=\int\alpha_\nu cd\eta=\hat{\kappa}_l\rho_lc\eta_l \;, \label{tau_l0}
\end{eqnarray}
where
\begin{eqnarray}
  \hat{\kappa}_l\equiv\frac{\pi e^2}{m_ec}\frac{f_{12}}{\nu_l}\frac{n_{1l}}{\rho_l}\left(1-e^{-h\nu_l/kT_l}\right)  \label{hkap_l}
\end{eqnarray}
is a scale parameter for the line opacity, $\rho_l\equiv\rho(\eta=\eta_l)$, $n_{1l}\equiv n_1(\eta=\eta_l)$, and $T_l\equiv T(\eta=\eta_l)$. In the evaluation of the integral we have adopted the convention that $\phi(\nu)=\delta(\nu-\nu_l)$, i.e., the line profile function is a Dirac $\delta$-function of frequency centered at $\nu_l$.

The mass density evolves with time according to $\rho\propto\eta^{-3}$. Then, since $\nu\propto\eta^{-1}$, we have $\rho_lc\eta_l=\rho_0c\eta_0(\nu_l/\nu_0)^2$. Hence, we can write the line optical depth as
\begin{eqnarray}
  \tau_l(\nu_0)=s\frac{\hat{\kappa}_l}{\kappa_\es}\frac{\nu_l^2}{\nu_0^2}\vartheta(\nu_0-\nu_l) \;, \label{tau_l}
\end{eqnarray}
where $\kappa_\es$ is the constant opacity of electron scattering, $s\equiv\kappa_\es\rho_0c\eta_0$, and the Heaviside step function $\vartheta(x)=1$ if $x\ge 0$, $0$ if $x<0$. The parameter $s$ is related to the optical thickness of the sphere to electron scattering. The optical depth from the center of the sphere to its surface due to electron scattering is related to the $s$ parameter by
\begin{eqnarray}
  \tau_\es=\frac{s\beta_R}{1+\beta_R} \;. \label{tau_es_beta}
\end{eqnarray}
For the Newtonian case with $\beta_R\ll 1$, we have $\tau_\es\approx s\beta_R$. For the extremely relativistic case with $\beta_R\approx 1$, we have $\tau_\es\approx s/2$.

In \citet{kar77} the $s$ parameter is called the expansion parameter, since when $s$ is large it is equal to the reciprocal of the relative Doppler shift of a photon between electron scatterings. Here we see that $s$ is more relevant to the optical depth of the sphere to electron scattering.

If the frequency interval defined from $\nu_0$ to certain $\nu_\eta\equiv\nu(\eta)<\nu_0$ contains a number of line frequencies, the total optical depth of the lines contained in the interval to the photon of energy $\nu_0$ is
\begin{eqnarray}
  \sum_l\tau_l(\nu_0)=s\sum_{\nu_\eta<\nu_l<\nu_0}\frac{\hat{\kappa}_l}{\kappa_\es}\frac{\nu_l^2}{\nu_0^2} \;. \label{tau_bb}
\end{eqnarray}

\section{Derivation of the Line Expansion Opacity}
\label{l_exp_opc}

In a general situation, including the relativistic case when the medium where photons propagate is expanding so that the photon frequency changes along the ray path, the radiative transfer equation takes the form of \citep{nov73}
\begin{eqnarray}
  \frac{d}{dl}\left(\frac{I_\nu}{\nu^3}\right)=\frac{\rho}{4\pi}\frac{\epsilon_\nu}{\nu^3}-\rho(\kappa_\nu+\kappa_s)\frac{I_\nu}{\nu^3} \;, \label{trf_eq0}
\end{eqnarray}
where $I_\nu$ is the specific intensity of radiation, $l$ the proper spatial distance along the null light ray as measured in a local rest frame of the medium, $\epsilon_\nu$ the specific emissivity, $\kappa_\nu=\alpha_\nu/\rho$ the absorption opacity, and $\kappa_s$ the scattering opacity. The expansion of equation (\ref{trf_eq0}) leads to
\begin{eqnarray}
  \frac{dI_\nu}{dl}=\rho\kappa_\nu S_\nu-\rho(\kappa_\nu+\kappa_s+\kappa_H)I_\nu \;, \label{rad_tr}
\end{eqnarray}
where the source function $S_\nu\equiv\epsilon_\nu/4\pi\kappa_\nu$, and
\begin{eqnarray}
  \kappa_H\equiv -\frac{3}{\rho\nu}\frac{d\nu}{dl} \;. \label{kap_H}
\end{eqnarray}

In a static medium the frequency of a photon does not vary along its path unless the photon is scattered inelastically, then $\kappa_H=0$ and equation (\ref{rad_tr}) reduces to the standard radiative transfer equation that we usually see in the literature \citep[see, e.g.,][]{cha60,arm72,mih78,ryb04}.

From equation (\ref{rad_tr}) we see that $\kappa_H$ plays the role of an effective absorption opacity. Since $\nu^{-1}d\nu/dl$ is purely determined by the kinematics of the medium \citep[the expansion, shear, and acceleration of the fluid matter, see][]{nov73}, this effective opacity is independent of the photon frequency. Hence, we can call it a universal expansion opacity. However, the universal expansion opacity is not related to true absorption of photons. It is only related to the change of the photon intensity caused by the motion of the matter. In the case of a uniformly expanding sphere, the $\kappa_H$ only causes dilution of the photon intensity as the matter expands.

In a uniformly expanding sphere we have $dl=cd\eta$ and $d\nu/dl=-\nu/c\eta$. Then, we have
\begin{eqnarray}
  \kappa_H=\frac{3}{\rho c\eta}=\frac{3\kappa_\es}{s}\left(\frac{\eta}{\eta_0}\right)^2 \;.
\end{eqnarray}
Thus, we must have $\kappa_H>\kappa_\es$ when $\eta>\eta_0(s/3)^{1/2}$. This fact indicates that $\kappa_H$ cannot always be ignored.

Similar to \citet{kar77}, we can define a mean free path associated with an absorption and/or scattering process by
\begin{eqnarray}
  \bar{X}=\int_0^\infty e^{-\tau_\eff(l)}dl \;, \hspace{1cm} \tau_\eff(l)=\int_0^l\kappa\rho dl \;, \label{bl_int}
\end{eqnarray}
where the integrals are defined along the light ray. However, in our definition of $\bar{X}$, for the reason to be given below, the contribution of the universal expansion opacity $\kappa_H$ is included (hence the subscript ``eff'' in $\tau_\eff$). That is, we have $\tau_\eff=\tau_\nu+\tau_s+\tau_H$. By $\tau_H=-\int(3/\nu)(d\nu/dl)dl=3\ln(\nu_0/\nu)$, we get
\begin{eqnarray}
  \bar{X}=\int_0^\infty e^{-(\tau_\nu+\tau_s)}\left(\frac{\nu}{\nu_0}\right)^3 dl \;. \label{l_bar}
\end{eqnarray}

In the definition of the mean free path in \citet{kar77}, the $\tau_H$ is not included. This is not a problem for their calculations where they have assumed that $\nu\approx\nu_l$ always. However, in our work it is necessary to include the contribution of $\kappa_H$, since otherwise the integral in the definition of $\bar{X}$ will diverge if the physical optical depth $\tau_\nu+\tau_s$ does not approach infinity as $l\rightarrow\infty$. In fact, as we will see in Section~\ref{case_es}, for the case of electron scattering we have $\tau_\es\rightarrow s/2$ as $l\rightarrow\infty$.

Since $\rho\propto\nu^3\propto\eta^{-3}$, equation (\ref{l_bar}) can also be written as
\begin{eqnarray}
  \bar{X}=\frac{1}{\rho_0}\int_0^{\tau_{\max}}\frac{1}{\kappa}e^{-\tau}d\tau \;,
\end{eqnarray}
where $\tau\equiv\tau_\nu+\tau_s$, and $\tau_{\max}\equiv\int_0^\infty\kappa\rho dl$. When $\kappa=\mbox{const}$, we get $\bar{X}=(\kappa\rho_0)^{-1}\left(1-e^{-\tau_{\max}}\right)$, which leads to $\kappa=\left(\rho_0\bar{X}\right)^{-1}\left(1-e^{-\tau_{\max}}\right)$. This relation motivates us to define an averaged opacity through the mean free path by
\begin{eqnarray}
  \bar{\kappa}=\frac{1}{\rho_0\bar{X}}\left(1-e^{-\tau_{\max}}\right) \;. \label{bar_kap}
\end{eqnarray}

The averaged opacity defined above is a function of the photon frequency at time $\eta=\eta_0$, i.e., a function of $\nu_0$.

\subsection{The case of electron scattering}
\label{case_es}

\begin{figure}[ht!]
\centering{\includegraphics[angle=0,scale=0.75]{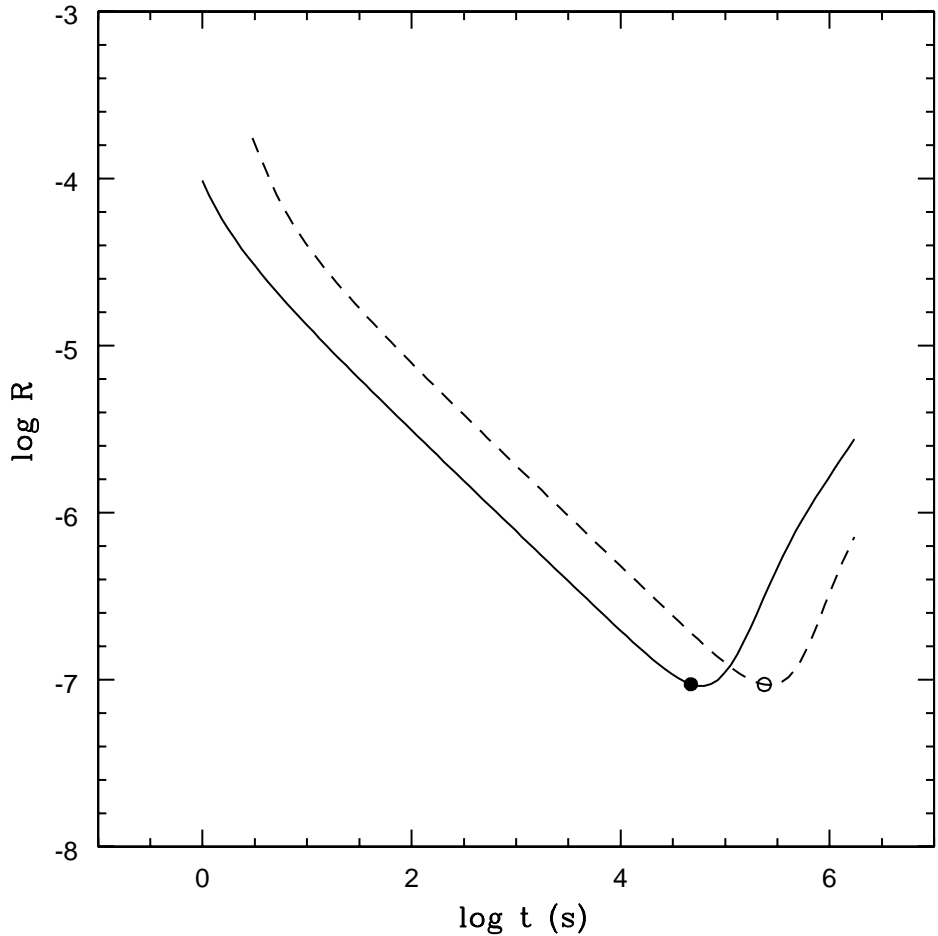}}
\caption{Variation of the parameter $R=\rho/T_6^3$ with time in the merger model fitting the kilonova emission of GW170817/ GRB170817A. The time $t$ is measured from the moment of merger of two neutron stars. The model consists of two ejecta components A and B as described in the Fig.~1 and Table~1 of \citet{li19}. The $R$ for component A is shown with the solid curve, where the dark point at $t\approx 0.5\,\oday$ denotes the time at the peak of the luminosity of component A. The $R$ for component B is shown with the dashed curve, where the circle at $t\approx 2.7\,\oday$ denotes the time at the peak of the luminosity of component B.
}
\label{R}
\end{figure}

When the mass density is low and the temperature is high, the continuous opacity in a matter is dominantly due to electron scattering \citep{igl96,sea04,car17}. This condition is evidently satisfied for supernovae and neutron star mergers near the peak of their luminosities. In Fig.~\ref{R} we plot the parameter $R=\rho/T_6^3$ against time for the merger model adopted by \citet{li19} to fit the kilonova emission associated with GW170817/GRB170817A, where the mass density of the merger ejecta $\rho$ is in $\g\,\cm^{-3}$, and the temperature $T$ is in $10^6\,\K$. The parameter $R$ is often used to characterize the Rosseland mean opacity in a plasma gas \citep{igl96,sea04}.

From Fig.~\ref{R} we see that, for both the ejecta components in the model, the value of $R$ is $\sim 10^{-7}$ near the peak of the luminosity. For component A (the solid curve) which has an expansion velocity of $0.3c$, at the time of the peak luminosity ($\approx 0.5\,\oday$ after the merger) we have $\rho\approx 1.6\times 10^{-13}\g\,\cm^{-3}$ and $T\approx 1.2\times 10^4\,\K$. For component B (the dashed curve) which has an expansion velocity of $0.1c$, at the time of the peak luminosity  ($\approx 2.7\,\oday$ after the merger) we have $\rho\approx 0.5\times 10^{-13}\g\,\cm^{-3}$ and $T\approx 0.8\times 10^4\,\K$.  These numbers indicate that near the peak of the luminosity the continuous opacity in the ejecta of the neutron star merger is dominantly contributed by electron scattering. The situation is similar for supernovae, although they may have a chemical composition very different from that of a neutron star merger.

Since the opacity of electron scattering plays a very important role in many astronomical phenomena including supernovae and neutron star mergers as we have just discussed, in this subsection we consider the opacity of electron scattering and calculate the corresponding mean free path.

The differential cross section for a free electron scattering unpolarized radiation derived in quantum electrodynamics, which has taken into account the effect of special relativity and quantum theory of electrons and radiation, is given by the Klein-Nishina formula \citep{kle29,hei54}. For low energy photons with $h\nu\ll m_ec^2$, the transfer of energy between the radiation and the electron is negligible and the resultant elastic scattering process is just the classical Thomson scattering. In this case, the Klein-Nishina cross section reduces to the classical Thomson cross section, which is a constant after integration over all directions. When $h\nu\ga m_ec^2$, the energy transfer between the photon and the electron is nonnegligible and then the resultant inelastic scattering process is called the Compton scattering where the relativistic and quantum effects must be taken into account.

In this paper we assume that the energy of a photon always satisfies the low energy condition $h\nu\ll m_ec^2$. Then, the cross section for electron scattering is simply given by the constant Thompson cross section, which leads to a constant continuous opacity if the fraction of free electrons in the matter does not evolve with time. We remark that the condition $h\nu\ll m_ec^2$ is satisfied most of time for the case of supernovae and neutron star mergers. For instance, for the merger model shown in Fig.~\ref{R}, when $t>1\,\s$ the temperature inside the ejecta of component A satisfies $T<5.6\times 10^7\,\K$, and when $t>3\,\s$ the temperature inside the ejecta of component B satisfies $T<5.1\times 10^7\,\K$. A constant opacity for electron scattering sets a convenient scale for measuring the amplitude of the line expansion opacity. The assumption of a constant opacity for electron scattering also allows us to solve the corresponding equations for electron scattering analytically.

With the assumption of a constant opacity $\kappa_\es$, the optical depth for electron scattering in a uniformly expanding sphere is given by
\begin{eqnarray}
  \tau_\es=\frac{s}{2}\left(1-\frac{\eta_0^2}{\eta^2}\right) \;, \label{tau_es2}
\end{eqnarray}
where it has been assumed that the light ray along which the optical depth is calculated starts from the center of the sphere at time $\eta_0$ and reaches a point off the center at time $\eta$. By $dl=cd\eta$ and $\nu/\nu_0=\eta_0/\eta$, we get the mean free path
\begin{eqnarray}
  \bar{X}=\int_0^\infty e^{-{\tau_\es}}\left(\frac{\nu}{\nu_0}\right)^3 dl=\frac{c\eta_0}{s}\left(1-e^{-s/2}\right) \;, \label{mfp_es}
\end{eqnarray}
when there is only electron scattering and no absorption.

The mean free path calculated here, $\bar{X}$, includes the contribution of the universal expansion opacity through the factor $(\nu/\nu_0)^3$ in the integral. The limit $s\rightarrow 0$ leads to the mean free path for the universal expansion, which is $X_H=c\eta_0/2$. By equation (\ref{tau_es2}), we get $\tau_{\max}=s/2$. Then, by equations (\ref{bar_kap}) and (\ref{mfp_es}) we get $\bar{\kappa}_\es=\kappa_\es$, which is just what we have expected.

\subsection{The mean free path and averaged opacity for spectral lines}
\label{case_lexp}

When there are both line absorption and electron scattering, the mean free path should be calculated by equation (\ref{l_bar}) with $\tau_\nu$ replaced by the $\sum_l\tau_l$ in equation (\ref{tau_bb}) and $\tau_s$ replaced by the $\tau_\es$ in equation (\ref{tau_es2}). Then, we get
\begin{eqnarray}
  \bar{X} = \int_0^\infty\exp\left(-s\sum_{l=J}^{N(\eta)}\frac{\hat{\kappa}_l}{\kappa_\es}\frac{\nu_l^2}{\nu_0^2}\right)\exp\left[-\frac{s}{2}\left(1-\frac{\eta_0^2}{\eta^2}\right)\right]\left(\frac{\nu}{\nu_0}\right)^3dl \;,
\end{eqnarray}
where the $J$-th line is the first one in the line list with a frequency smaller than $\nu_0$ (i.e., the first line encountered by a photon of frequency $\nu_0$), and the $N(\eta)$-th is the last line approached by the photon frequency by the time $\eta$.

Since $\eta_0/\eta=\nu/\nu_0$ and $dl=cd\eta=-c\eta_0\nu_0d\nu/\nu^2$, we get
\begin{eqnarray}
  \bar{X} = \frac{1}{2}c\eta_0e^{-s/2} \int_0^1\exp\left(-s\sum_{l=J}^{N(y)}\frac{\hat{\kappa}_l}{\kappa_\es}\frac{\nu_l^2}{\nu_0^2}\right)e^{sy/2}dy \;,
\end{eqnarray}
where $y\equiv\nu^2/\nu_0^2$, $N(y)=N(\eta=\eta_0 y^{-1/2})$, and we have used the fact that $\nu=0$ as $l\rightarrow\infty$.

Following \citet{kar77}, we divide the $y$-coordinate into a series of segments: $[1, y_J+\epsilon]$, $[y_J+\epsilon, y_{J+1}+\epsilon]$, ... $[y_N+\epsilon, 0]$, where $\epsilon$ is a positive infinitesimal number, $y_J=\nu_J^2/\nu_0^2$, etc., and $N=N(\eta\rightarrow\infty)$ corresponds to the last line in the line list. Each frequency segment contains a line, except the first segment $[1, y_J+\epsilon]$ which contains no line. After the integration done on each segment, we get
\begin{eqnarray}
  \bar{X}=\frac{c\eta_0}{s}\left\{1-e^{-s(1-y_J)/2}+\sum_{j=J}^N\exp\left(-s\sum_{l=J}^j\frac{\hat{\kappa}_l}{\kappa_\es}\frac{\nu_l^2}{\nu_0^2}\right)\left[e^{-s(1-y_j)/2}-e^{-s(1-y_{j+1})/2}\right]\right\} \;, \label{mfp1}
\end{eqnarray}
where $y_j=\nu_j^2/\nu_0^2$ and $y_{N+1}=\nu^2_\infty/\nu_0^2=0$.

After reordering the terms in the sum in equation (\ref{mfp1}), we get
\begin{eqnarray}
  \bar{X}=X_\es\left\{1-\exp\left(-\sum_{i=J}^N\tau_i-\frac{s}{2}\right)-\sum_{j=J}^N\left(1-e^{-\tau_j}\right)\exp\left[-\sum_{i=J}^{j-1}\tau_i-\frac{s}{2}(1-y_j)\right]\right\} \;, \label{mfp3}
\end{eqnarray}
where we have written $c\eta_0/s=(\kappa_\es\rho_0)^{-1}\equiv X_\es$ (a scale for the mean free path of electron scattering), and
\begin{eqnarray}
  \tau_i\equiv s\frac{\hat{\kappa}_i}{\kappa_\es}\frac{\nu_i^2}{\nu_0^2}  \label{tau_j}
\end{eqnarray}
is the optical depth of the $i$-th line (see eq.~\ref{tau_l}). In derivation of equation~(\ref{mfp3}) we have made use of the convention that $\sum_{i=J}^{J-1}\tau_i=0$.

It is easy to check that when all $\tau_i=0$, the mean free path given by equation (\ref{mfp3}) returns to that for electron scattering in equation (\ref{mfp_es}).

The maximum optical depth is $\tau_{\max}=\sum_{i=J}^N\tau_i+s/2$. Then, from equations (\ref{bar_kap}) and (\ref{mfp3}) we get the line expansion opacity
\begin{eqnarray}
  \kappa_\expp(\nu_0) =\frac{\kappa_\es}{1-w(\nu_0)} \;, \label{kap_exp1a}
\end{eqnarray}
where as in \citet{kar77} we have introduced an enhancement factor
\begin{eqnarray}
  w(\nu_0)={\cal C}_J^{-1}\sum_{j=J}^N\left(1-e^{-\tau_j}\right)\exp\left[-\sum_{i=J}^{j-1}\tau_i-\frac{s}{2}(1-y_j)\right] \;, \label{kap_exp1b}
\end{eqnarray}
where the factor ${\cal C}_J$ is defined by
\begin{eqnarray}
  {\cal C}_J\equiv 1-e^{-\tau_{\max}}=1-\exp\left(-\sum_{i=J}^N\tau_i-\frac{s}{2}\right) \;. \label{cal_CJ}
\end{eqnarray}

Equations (\ref{mfp3}) and (\ref{kap_exp1a})--(\ref{cal_CJ}) are among the most important results of this paper, which determine the mean free path of a photon in a uniformly expanding sphere with electron scattering and line absorption, and the evaluation of the corresponding line expansion opacity.

\section{Discussion on the Line Expansion Opacity}
\label{discuss1}

\subsection{Comparison to the result of \citet{kar77}}

Comparison of equation~(\ref{mfp3}) to the equation~(12) of \citet{kar77} tells us two differences in the mean free path derived in this paper and that derived in \citet{kar77}. First, the mean free path in equation~(\ref{mfp3}) contains a term $e^{-\tau_{\max}}=\exp\left(-\sum\tau_i-s/2\right)$ in the braces, which is absent in the equation~(12) of \citet{kar77}. This is caused by the fact that in the derivation of \citet{kar77} it has been assumed that the total optical depth of electron scattering from the center of the matter to infinity is infinite, which is not valid in the case of a uniformly expanding sphere as we have shown in Section~\ref{case_es}. When the matter is optically thick, we have $e^{-\tau_{\max}}\approx 0$ so the term is not important. However, when the matter is optically thin, the term will be important.

Second, in our result the contribution of electron scattering to the integral of the mean free path in each frequency segment is represented by the term
\begin{eqnarray}
  \exp\left[-\frac{s}{2}(1-y_j)\right]=\exp\left[-\frac{s}{2}\left(1-\frac{\nu_j^2}{\nu_0^2}\right)\right] \;,
\end{eqnarray}
while in the equation~(12) of \citet{kar77} the corresponding contribution is represented by
\begin{eqnarray}
  \exp\left(-\kappa_\es\rho x_j\right)=\left(\frac{\nu_j}{\nu_0}\right)^s \;.
\end{eqnarray}
In the model of \citet{kar77} it has been assumed that $\nu_j$ is always close to $\nu_0$ and the mass density does not vary as the photon travels a distance of the mean free path, which is possible only in the diffusion limit with $s\gg 1$. Let us write $\nu_j=\nu_0(1-\delta)$, $0<\delta\ll 1$. Then, we have $\exp\left[-s(1-y_j)/2\right]\approx\exp(-s\delta)$ and $(\nu_j/\nu_0)^s\approx\exp(-s\delta)$. Hence, in the diffusion limit our result agrees with that of \citet{kar77} (see Sec.~\ref{large_s}). However, in the optically thin case with $s\ll 1$, or in the transitional case when $s$ is $\sim 1$, our formula gives more accurate results.

The same conclusions are obtained if we compare the line expansion opacity in equations (\ref{kap_exp1a})--(\ref{cal_CJ}) to the equation~(13) of \citet{kar77}.

\subsection{The case without electron scattering}
\label{small_s}

To get the solution for the case without electron scattering (i.e., for the case of pure line absorption), we need to take the limit $\kappa_\es\rightarrow 0$, i.e., $s\rightarrow 0$ but $s/\kappa_\es$ remains finite. For small $s$, we have $\exp[-(s/2)(1-y_j)]\approx 1-(s/2)(1-y_j)$. Then, making use of the identity
\begin{eqnarray}
  \sum_{j=J}^N\left(1-e^{-\tau_j}\right)\exp\left(-\sum_{i=J}^{j-1}\tau_i\right)=1-\exp\left(-\sum_{i=J}^N\tau_i\right) \;, \label{sum_id}
\end{eqnarray}
as $\kappa_\es\rightarrow 0$ we get
\begin{eqnarray}
  \bar{X}=\frac{c\eta_0}{2}\left[1-\sum_{j=J}^N\left(1-e^{-\tau_j}\right)\exp\left(-\sum_{i=J}^{j-1}\tau_i\right)y_j\right] \;. \label{mfp4} 
\end{eqnarray}

When all $\tau_j=0$, we get $\bar{X}=c\eta_0/2$, which is just the mean free path arising from the universal expansion opacity.

The line expansion opacity derived from equations (\ref{bar_kap}) and (\ref{mfp4}) is
\begin{eqnarray}
  \kappa_\expp=\frac{2}{\rho_0c\eta_0}{\cal C}_{J,0}\left[1-\sum_{j=J}^N\left(1-e^{-\tau_j}\right)\exp\left(-\sum_{i=J}^{j-1}\tau_i\right)y_j\right]^{-1} \;,
\end{eqnarray}
where ${\cal C}_{J,0}={\cal C}_J(s=0)=1-\exp\left(-\sum_{i=J}^N\tau_i\right)$.

When all $\tau_j=0$, we get $\kappa_\expp=0$ as we would have expected.

\subsection{The case of optically thick to electron scattering}
\label{large_s}

Here by ``optically thick to electron scattering'' we mean that $\tau_\es\gg 1$, which automatically implies $s\gg 1$ by equation (\ref{tau_es_beta}). In this limit, equation (\ref{mfp3}) becomes
\begin{eqnarray}
  \bar{X}\approx X_\es\left\{1-\sum_{j=J}^N\left(1-e^{-\tau_j}\right)\exp\left[-\sum_{i=J}^{j-1}\tau_i-\frac{s}{2}(1-y_j)\right]\right\} \;,
\end{eqnarray}
where $y_j=\nu_j^2/\nu_0^2<1$. From this equation we see that when $s\gg 1$, the line contribution to the opacity is important only if $y_j\approx1$, i.e., $\nu_j\approx\nu_0$. Then, we have $1-y_j=(\nu_0+\nu_j)(\nu_0-\nu_j)/\nu_0^2\approx 2(1-\nu_j/\nu_0)$. Since $\ln(\nu_0/\nu_j)\approx 1-\nu_j/\nu_0$ when $0<1-\nu_j/\nu_0\ll 1$, we have $1-y_j\approx 2\ln(\nu_0/\nu_j)$.

Hence, in the limit $s\gg 1$ we get
\begin{eqnarray}
  \bar{X}\approx X_\es\left[1-\sum_{j=J}^N\left(1-e^{-\tau_j}\right)\left(\frac{\nu_j}{\nu_0}\right)^s\exp\left(-\sum_{i=J}^{j-1}\tau_i\right)\right] \;, \label{mfp5}
\end{eqnarray}
and
\begin{eqnarray}
  \kappa_\expp\approx\kappa_\es\left[1-\sum_{j=J}^N\left(1-e^{-\tau_j}\right)\left(\frac{\nu_j}{\nu_0}\right)^s\exp\left(-\sum_{i=J}^{j-1}\tau_i\right)\right]^{-1} \;. \label{kap_exp3}
\end{eqnarray}

Equation (\ref{kap_exp3}) agrees with the equation (13) of \citet{kar77}. Thus, in the limit $s\rightarrow\infty$ our result approaches that of \citet{kar77}. We can also claim that the formulae derived by \citet{kar77} are accurate only in the case of $s\gg 1$.

\subsection{The limit of high density weak lines}

Consider a group of absorption lines with frequency in the range of $\nu_0$ to $\nu_0-\Delta\nu$, $\Delta\nu\ll\nu_0$. The lines have equal frequency spacing $\delta\nu$, and equal optical thickness $\tau$. The first line is the $J$-th, and the last one is the $N^\prime$-th, $N^\prime<N$. Then, the total number of lines in the given frequency interval is $\hat{N}=N^\prime-J+1$. The frequency of the $i$-th line is then $\nu_i=\nu_0-(i-J+1)\delta\nu$, and we have $\Delta\nu=(N^\prime-J+1)\delta\nu$.

We assume that the distribution of the lines in the frequency space is sufficiently dense, so that the total line optical depth within the small frequency interval $\Delta\nu$ is large: $\hat{N}\tau\gg 1$. Then, in the sum $\sum_{j=J}^N$ in equation (\ref{mfp3}), all terms with $j>N^\prime$ can be ignored because of the factor $\exp\left(-\sum_{i=J}^{j-1}\tau_i\right)$. Therefore, in the dense line approximation, the sum $\sum_{j=J}^N$ can be replaced by $\sum_{j=J}^{N^\prime}$, and we can take all $\nu_j\approx \nu_0$ in equation (\ref{mfp3}).

Based on the above description, we have $\sum_{i=J}^{j-1}\tau_i=(j-J)\tau$, $\sum_{i=J}^{N^\prime}\tau_i=\left(N^\prime-J+1\right)\tau=\hat{N}\tau$, and $y_j=[1-(i-J+1)\delta\nu/\nu_0]^2\approx 1-2(j-J+1)\delta\nu/\nu_0$. Then, in the dense line approximation equation (\ref{mfp3}) leads to
\begin{eqnarray}
  \bar{X} \approx X_\es\left\{1-\left(e^{\tau}-1\right)\sum_{n=1}^{\hat{N}}\exp\left[-n\left(\tau+s\frac{\delta\nu}{\nu_0}\right)\right]\right\} \;,
\end{eqnarray}
where $n=j-J+1$. By the identity $\sum_{n=1}^{\hat{N}}e^{-na}=(e^a-1)^{-1}\left[1-\exp\left(-\hat{N}a\right)\right]$, we get
\begin{eqnarray}
  \bar{X} \approx X_\es\left\{1-\left(e^{\tau}-1\right)\frac{1-\exp\left[-\hat{N}(\tau+s\delta\nu/\nu_0)\right]}{\exp\left(\tau+s\delta\nu/\nu_0\right)-1}\right\} \approx X_\es\left[1-\frac{e^\tau-1}{\exp\left(\tau+s\delta\nu/\nu_0\right)-1}\right] \;. \label{barl_exp}
\end{eqnarray}

Now let us consider the case of $\tau\ll 1$ but $\hat{N}\tau\gg 1$, i.e., the case of high density weak lines. Then, we have $e^\tau\approx1+\tau$ and $e^{-\hat{N}\tau}\approx 0$, and by equation~(\ref{barl_exp}) we get
\begin{eqnarray}
  \bar{X} \approx X_\es\left[1-\frac{\tau}{\exp\left(\tau+s\delta\nu/\nu_0\right)-1}\right] \;.
\end{eqnarray}
We further assume that $s\delta\nu/\nu_0\ll 1$, i.e., $s\ll \nu_0/\delta\nu$, which means that a photon encounters many lines between electron scatterings. Then, we have $\exp\left(\tau+s\delta\nu/\nu_0\right)-1\approx \tau+s\delta\nu/\nu_0$, and
\begin{eqnarray}
  \bar{X}\approx X_\es\frac{1}{1+\tau\nu_0/s\delta\nu} \;. \label{l_den}
\end{eqnarray}

From equation (\ref{l_den}) we derive that
\begin{eqnarray}
  \kappa_\expp=(\rho_0\bar{X})^{-1}\approx \kappa_\es\left(1+\frac{\tau\nu_0}{s\delta\nu}\right) \;,
\end{eqnarray}
where we have used the fact that $1-e^{-\tau_{\max}}\approx 1$. Thus, in the case of high density weak lines, we have the line expansion opacity
\begin{eqnarray}
  \kappa_{\expp,{\rm line}}\approx\kappa_\es\frac{\tau\nu_0}{s\delta\nu} =\frac{1}{\rho_0c\eta_0} \frac{\nu_0}{\Delta\nu}\hat{N}\tau \;. \label{kap_w}
\end{eqnarray}

Therefore, even though the optical depth of each line is small, the line expansion opacity can be considerably greater than the conventional opacity as a result of cumulative effect of many weak lines if $(s/\tau)(\delta\nu/\nu_0) \ll 1$, as having been demonstrated by \cite{kar77}.

\subsection{The case of strong thick lines}

In the dense line approximation, when $\tau\gg 1$ from equation~(\ref{barl_exp}) we have $\bar{X}\approx X_\es\left(1-e^{-s\delta\nu/\nu_0}\right)$. Then, we have the expansion opacity
\begin{eqnarray}
  \kappa_\expp\approx \kappa_\es\left(1-e^{-s\delta\nu/\nu_0}\right)^{-1} \;,
\end{eqnarray}
and the line expansion opacity
\begin{eqnarray}
  \kappa_{\expp,{\rm line}}=\kappa_\expp-\kappa_\es\approx \kappa_\es\left(e^{s\delta\nu/\nu_0}-1\right)^{-1} \;. \label{kap_s0}
\end{eqnarray}

If we further assume that $s\delta\nu/\nu_0\ll 1$, i.e., no electron scattering occurs before a photon is absorbed by a line, we have $e^{s\delta\nu/\nu_0}-1\approx s\delta\nu/\nu_0$ and
\begin{eqnarray}
  \kappa_{\expp,{\rm line}}\approx\frac{\kappa_\es}{s}\frac{\nu_0}{\delta\nu}=\frac{1}{\rho_0c\eta_0} \frac{\nu_0}{\Delta\nu}\hat{N} \;. \label{kap_s}
\end{eqnarray}

When $\tau\gg 1$, the mean free path is essentially equal to the distance traveled by a photon from one line to the next. Since $\delta\eta=\eta_0\delta\nu/\nu_0$, we have $X_{\rm line}=c\delta\eta=c\eta_0\delta\nu/\nu_0$. Then we get $\kappa_{\rm line}=(\rho_0X_{\rm line})^{-1}=(\rho_0c\eta_0)^{-1}\nu_0/\delta\nu$, which is just the line expansion opacity given by equation (\ref{kap_s}).

The solutions in equations (\ref{kap_w}) and (\ref{kap_s}) can be combined to form a general solution for the case of dense lines
\begin{eqnarray}
  \kappa_{\expp,{\rm line}}=\frac{1}{\rho_0c\eta_0} \frac{\nu_0}{\delta\nu}\left(1-e^{-\tau}\right) \;, \label{kap_ws}
\end{eqnarray}
which applies to the case of any value of $\tau$. However, the dense line condition $s\delta\nu/\nu_0\ll 1$ must be satisfied.

From the above results we find that $\kappa_{\expp,{\rm line}}\propto\hat{N}/\Delta\nu=1/\delta\nu$, no matter whether $\tau$ is large or small. Therefore, the line expansion opacity is proportional to the number of lines in a fixed frequency interval, i.e., proportional to the density of lines on the wavelength coordinate.

In numerical works for calculation of the expansion opacity arising from bound-bound transitions of heavy atoms in supernovae and neutron star mergers, people usually use the following simple formula to save the computing efficiency \citep[see, e.g.,][]{kas13,tan13,gai19,tan19}:
\begin{eqnarray}
  \kappa_\expp=\frac{1}{\rho c t}\sum_i\frac{\lambda_i}{\Delta\lambda}\left(1-e^{-\tau_i}\right) \;, \label{simp_kap}
\end{eqnarray}
where $\lambda_i$'s are the wavelength of the lines contained in a small wavelength interval $\Delta\lambda$. This simplified expression for the expansion opacity was first introduced by \citet{eas93}. It agrees with our equation~(\ref{kap_ws}), considering the fact that $\lambda=c/\nu$, and $t=\eta$ in the Newtonian limit.

Since equation~(\ref{kap_ws}) was derived from the mean free path in equation (\ref{mfp3}) under the assumption that the spectral lines are densely and uniformly distributed in a small frequency interval, the validity of equation~(\ref{simp_kap}) should also be restricted by the same assumption. When the assumption of a dense and uniform distribution of lines in the frequency/wavelength space is violated, the line expansion opacity calculated with equation~(\ref{simp_kap}) would contain large errors. An obvious problem with the formalism defined by equation~(\ref{simp_kap}) is that the value of $\kappa_\expp$ calculated with it depends on the choice of the size of $\Delta\lambda$ for sparsely and/or nonuniformly distributed spectral lines.

\section{The Rosseland Mean Opacity}
\label{ross}

The Rosseland mean opacity $\kappa_R$ is defined by
\begin{eqnarray}
  \kappa_R^{-1}\equiv\llangle\kappa^{-1}\rrangle_R\equiv \frac{\int_0^\infty\kappa^{-1}(\partial B_\nu/\partial T)d\nu}{\int_0^\infty(\partial B_\nu/\partial T)d\nu} \;, \label{ros_opc}
\end{eqnarray}
where $B_\nu=B_\nu(T)$ is the Planck function for blackbody radiation. Submitting equation (\ref{kap_exp1a}) into equation (\ref{ros_opc}), we get the Rosseland mean of the line expansion opacity\footnote{The Planck mean has been applied to the line expansion opacity in some references, e.g., in \citet{kas13,tan13,gai19,tan19}. Here we point out that the Planck mean is useful only in the optically thin case \citep{arm72,zel02}. When the medium is optically thick---which must be the case for the early stage of supernovae and neutron star mergers---the Rosseland mean is more appropriate \citep{arm72,nov73,zel02,ryb04}.}
\begin{eqnarray}
  \kappa_{\expp,R}=\frac{\kappa_\es}{1-\langle w\rangle_R} \;, \hspace{1cm} \langle w\rangle_R\equiv \frac{\int_0^\infty w(\nu) (\partial B_\nu/\partial T)d\nu}{\int_0^\infty(\partial B_\nu/\partial T)d\nu} \;. \label{ros_w}
\end{eqnarray}

Define $x\equiv h\nu/kT$, we get
\begin{eqnarray}
  \langle w\rangle_R =\frac{15}{4\pi^4}\int_0^\infty w(x)f_{dB}(x)dx \;, \hspace{1cm} f_{dB}(x)\equiv \frac{x^4e^x}{(e^x-1)^2} \;, \label{w_int}
\end{eqnarray}
where $w(x)\equiv w(\nu=kTx/h)$.

\begin{figure}[hb!]
  \vspace{0.3cm}
\centering{\includegraphics[angle=0,scale=0.85]{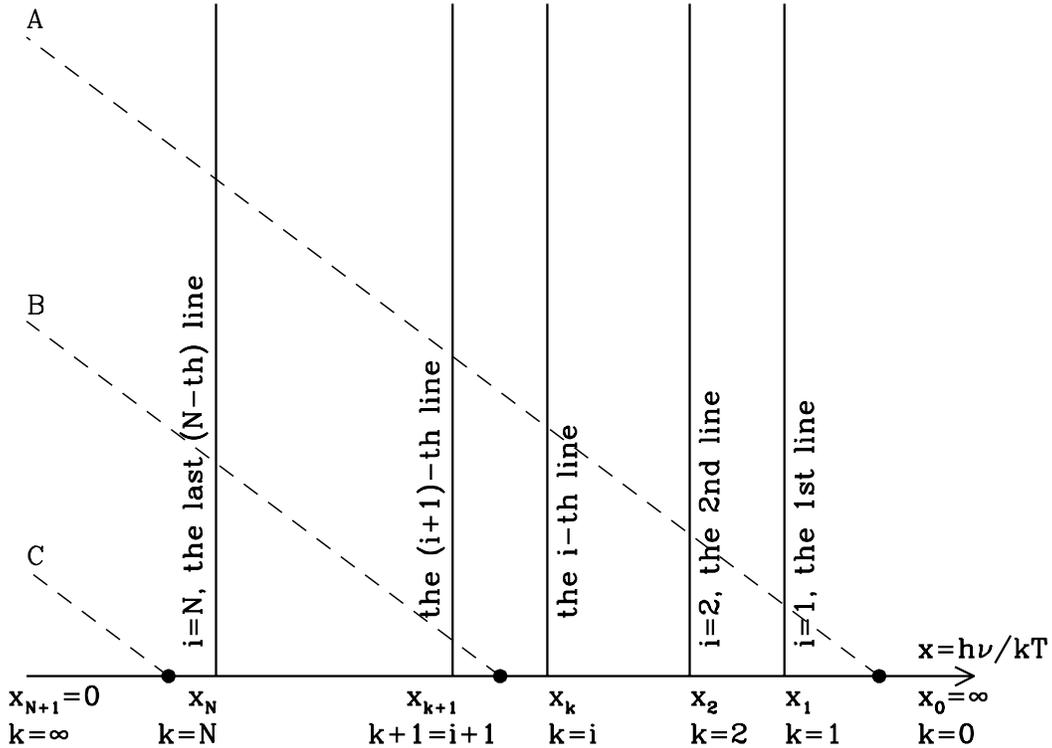}}
\caption{To evaluate the Rosseland mean of the line expansion opacity, the integral over $x=h\nu/kT$ is divided into segments defined by intervals $[x_{N+1}=0,x_N]$, $[x_N,x_{N-1}]$, ..., $[x_{k+1},x_k]$, ..., $[x_2,x_1]$, $[x_1,x_0=\infty]$. A photon with a frequency $\nu$ in the interval $[x_{k+1},x_k]$ will pass by all spectral lines with a frequency $\nu_i<\nu$, i.e., those lines from the $(J=k+1)$-th to the last one i.e. the $N$-th (dashed line B). Hence, a photon with a frequency in the interval $[x_1,x_0=\infty]$ will pass by all the lines (dashed line A; $J=1$), while a photon with a frequency in the interval $[x_{N+1}=0,x_N]$ will not pass by any line (dashed line C).
}
\label{ros}
\end{figure}

As in the derivation of the monochromatic expansion opacity, the integral over $x$ in equation (\ref{w_int}) can be broken up into segments that run from one line to the next, giving
\begin{eqnarray}
  \langle w\rangle_R =\frac{15}{4\pi^4}\sum_{k=0}^N\int_{x_{k+1}}^{x_k}w(x)f_{dB}(x)dx \;, \label{w_int2}
\end{eqnarray}
where $x_0=\infty$ and $x_{N+1}=0$. However, here each segment ends on two neighbor lines, unlike in the case in calculation of the monochromatic opacity where each segment ends on an intermediate point between two neighbor lines in order to evaluate the integral involving Dirac $\delta$-functions. For a photon frequency $\nu$ in the interval $[x_{k+1},x_k]$, as time goes on it will pass by all spectral lines with frequency $\nu_i<\nu$ from the first one labeled by the index $J=k+1$, to the last one, i.e., the $N$-th line (Fig.~\ref{ros}).

Since a photon in the interval of $[x=0,x_N]$ will not pass by any line, it only suffers the electron scattering and thus has $w=0$. This means that the segment defined by the interval of $[x=0,x_N]$ makes zero contribution to the evaluation of $\langle w\rangle_R$. Then, submitting equation (\ref{kap_exp1b}) into equation (\ref{w_int2}), we derive that
\begin{eqnarray}
  \langle w\rangle_R =\frac{15}{4\pi^4}\sum_{k=0}^{N-1}\sum_{j=J_k}^N\int_{x_{k+1}}^{x_k}{\cal C}_{J_k}^{-1}(x)\left[1-e^{-\tau_j(x)}\right]\exp\left\{-\sum_{i=J_k}^{j-1}\tau_i(x)-\frac{s}{2}\left[1-y_j(x)\right]\right\}f_{dB}(x)dx \;, \label{w_cal}
\end{eqnarray}
where $J_k=k+1$,
\begin{eqnarray}
  y_j(x)=\left(\frac{h\nu_j}{kT}\right)^2 x^{-2} \;, \hspace{1cm} \tau_j(x)=s\frac{\hat{\kappa}_j(x)}{\kappa_\es}y_j(x) \;, \label{y_tau_j}
\end{eqnarray}
and
\begin{eqnarray}
  \hat{\kappa}_j(x)=\frac{\pi e^2}{m_ec}\frac{f_{12,j}}{\nu_j}\frac{n_{1,j}}{\rho_j}\left(1-e^{-x}\right) \;. \label{h_kap_j}
\end{eqnarray}
Here for simplicity we have assumed that $T\propto\eta^{-1}\propto\nu$ and then $T_j=T\nu_j/\nu$, $h\nu_j/kT_j=h\nu/kT=x$.

The opacity of electron scattering is defined by $\kappa_\es=\sigma_\Th n_e/\rho$, where $n_e$ is the number density of free electrons, $\sigma_\Th=(8\pi/3)r_e^2$ is the Thomson cross-section, and $r_e=e^2/m_e c^2$ is the classical electron radius. Then, we have
\begin{eqnarray}
  \frac{\hat{\kappa}_j}{\kappa_\es}=\frac{3}{8}\frac{f_{12,j}\lambda_j}{r_e}\frac{n_{1,j}}{n_{e,j}}\left(1-e^{-x}\right) \sim \frac{f_{12,j}\lambda_j}{r_e}\frac{n_{1,j}}{n_{e,j}} \;, \label{h_kap_j2}
\end{eqnarray}
where $n_{e,j}=n_e(\eta=\eta_j)$.

If we define $n_j=n_{1,j}+n_{2,j}$, then by $n_{2,j}/n_{1,j}=(g_{2,j}/g_{1,j})e^{-h\nu_j/kT_j}=(g_{2,j}/g_{1,j})e^{-x}$ we get
\begin{eqnarray}
  \frac{n_{1,j}}{n_j}=\left(1+\frac{g_{2,j}}{g_{1,j}}e^{-x}\right)^{-1} \;.
\end{eqnarray}
Since $x=h\nu/kT>0$, we find that $\left(1+g_{2,j}/g_{1,j}\right)^{-1}<n_{1,j}/n_j<1$. For $g_{2,j}/g_{1,j}\sim 1$, we get $0.5<n_{1,j}/n_j<1$. Hence, we expect that $n_{1,j}/n_j$ is a slow function of $x$. We can also assume that $n_j/\rho_j$ is a slow function of $x$. Then, in the calculation of the Rosseland mean opacity for simplicity we can ignore the variation of $n_{1,j}/\rho_j$ with respect to the photon frequency.

Let us first discuss the solution of $\langle w\rangle_R$ (hence $\kappa_{\exp,R}$, which is a monotonically increasing function of $\langle w\rangle_R$) in the limit $s\rightarrow 0$ and $s\rightarrow\infty$. By the definition of $s$, if we fix the value of $\kappa_\es$ we have $s\propto\rho_0\eta_0\propto\eta_0^{-2}$. Hence, $s\rightarrow 0$ corresponds to the case of $\rho_0\rightarrow 0$ or $\eta_0\rightarrow\infty$ (i.e., the case of low mass density); and $s\rightarrow\infty$ corresponds to the case of $\rho_0\rightarrow\infty$ or $\eta_0\rightarrow 0$ (i.e., the case of high mass density). By the definition of $\tau_i$, we have $\tau_i\propto\rho_i\eta_i=\rho_0\eta_0(\nu_i/kT)^2x^{-2}$. Hence, when the value of $\kappa_\es$ is fixed, we have $\tau_i\propto s$. Then, $s\rightarrow 0$ implies $\tau_i\rightarrow 0$, and $s\rightarrow\infty$ implies $\tau_i\rightarrow\infty$.

Hence, as $s\rightarrow 0$, we have $\tau_i\rightarrow 0$ and $\tau_{\max}\rightarrow 0$. Then, we have
\begin{eqnarray}
  {\cal C}_J\approx \tau_{\max}\approx s\left(\sum_{i=J}^N\frac{\hat{\kappa}_i}{\kappa_\es}y_i+\frac{1}{2}\right) \;, \hspace{1cm} s\ll 1 \;.
\end{eqnarray}
By $1-e^{-\tau_j}\approx\tau_j$, from equation~(\ref{w_cal}) we get
\begin{eqnarray}
  \langle w\rangle_R = \frac{15}{4\pi^4}\sum_{k=0}^{N-1}\int_{x_{k+1}}^{x_k}\left(\sum_{i=J_k}^N\frac{\hat{\kappa}_i}{\kappa_\es}y_i+\frac{1}{2}\right)^{-1}\sum_{i=J_k}^N\frac{\hat{\kappa}_i}{\kappa_\es}y_if_{dB}(x)dx \;. \label{wR_s0}
\end{eqnarray}
Thus, as $s\rightarrow 0$ the Rosseland mean of the line expansion opacity approaches a finite and nonzero value. This conclusion differs from that in \citet{kar77}, where it was claimed that $\langle w\rangle_R\rightarrow 0$ as $s\rightarrow 0$. As we have discussed before, the expression for the line expansion opacity derived in \citet{kar77} does not apply to the case of small $s$.

From equation (\ref{w_cal}) we see that $\langle w\rangle_R \rightarrow 0$ as $s\rightarrow\infty$ and $\tau_i\rightarrow\infty$, since ${\cal C}_J(s\rightarrow\infty)=1$ and $1-y_j> 0$ always. As we have shown in Section~\ref{large_s}, as $s\rightarrow\infty$ our expression for the line expansion opacity approaches that in \citet{kar77}. Therefore, the limit of $s\rightarrow\infty$ derived above agrees with that derived in \citet{kar77}, i.e., $\langle w\rangle_R\approx{\cal O}(s^{-1})$. This result will also be proved in Section~\ref{apr}.

Next, let us discuss the variation of $\langle w\rangle_R$ with respect to the temperature $T$ for a given value of $s$.

\subsection{The case of $T\rightarrow 0$}
\label{asym1}

The function $f_{dB}(x)$ peaks at $x=x_m\equiv 3.83$, i.e., at $h\nu=h\nu_m\equiv 3.83kT$. For $x\gg x_m$, we have $f_{dB}(x)\approx x^4 e^{-x}$. For $x\ll x_m$, we have $f_{dB}(x)\approx x^2$. Hence, we expect that the dominant contribution to the integral of $\langle w\rangle_R$ comes from photon frequencies around $\nu_m\equiv 3.83kT/h$, i.e., around $x_m=3.83$.

As $T\rightarrow 0$, all lines have their frequencies $\nu_i\gg \nu_m$, and $x_i\equiv h\nu_i/kT\gg x_m$. Since in the expression of $\langle w\rangle_R$ in equation (\ref{w_cal}) all nonzero integrations come from frequency segments with $x>x_N\gg 1$, in the integral we can take $f_{dB}(x)\approx x^4 e^{-x}$. Because of the factor $e^{-x}$ and the fact that $x\gg 1$, the dominant contribution to the integral comes from the integrand evaluated near $x\approx x_N$.

The frequency separation between the $N$-th and the $(N-1)$-th line is $\nu_{N-1}-\nu_N$, which is fixed for a given sample of lines. As $T\rightarrow 0$, we must have $h(\nu_{N-1}-\nu_N)/kT\gg 1$, i.e., $x_{N-1}-x_N\gg 1$. In this limit, only the segment $[x_N,x_{N-1}]$ makes the dominant contribution to the integral.

When $x\approx x_N$, we have $1-y_N\approx(2/x_N)(x-x_N)$ and $\tau_N\approx (s/\kappa_\es)(\pi e^2/m_ec)(f_{12,N}/\nu_N)(n_{1,N}/\rho_N)$. Such a $\tau_N$ is independent of $x$, so is the ${\cal C}_N\approx 1-\exp(-\tau_N-s/2)$. Then from equation (\ref{w_cal}) we get
\begin{eqnarray}
  \langle w\rangle_R\approx\frac{15}{4\pi^4}{\cal C}_N^{-1}\left(1-e^{-\tau_N}\right)\int_{x_N}^{x_{N-1}}\exp\left[-\frac{s}{x_N}(x-x_N)\right]f_{dB}(x)dx \;,
\end{eqnarray}
where we have used the convention $\sum_{i=N}^{N-1}\tau_i=0$. By the approximation $f_{dB}(x)\approx x^4 e^{-x}$, after setting $x_{N-1}=\infty$ we get
\begin{eqnarray}
  \langle w\rangle_R \approx \frac{15}{4\pi^4}\frac{1-e^{-\tau_N}}{1-e^{-\tau_N-s/2}}\frac{x_N^4e^{-x_N}}{1+s/x_N}\zeta(s,x_N)  \;, \label{w_T_lim1a}
\end{eqnarray}
where $x_N=h\nu_N/kT$, $\tau_N$ is independent of $T$, and
\begin{eqnarray}
  \zeta(s,x_N)\equiv 1+4(x_N+s)^{-1}+12(x_N+s)^{-2}+24(x_N+s)^{-3}+24(x_N+s)^{-4} \;. \label{zeta_sxn}
\end{eqnarray}

Therefore, in the limit of $T\rightarrow 0$ (or, equivalently, $x_N\rightarrow\infty$), we have $\langle w\rangle_R \propto x_N^4e^{-x_N}$.

\subsection{The case of $T\rightarrow\infty$}
\label{asym2}

When $T\rightarrow\infty$, we have $\nu_m=3.83kT/h\gg \nu_i$ for all $1\le i\le N$. The dominant contribution to the integral of $\langle w\rangle_R$ comes from photons of frequency around $\nu_m$, which swipe all lines. In this limiting case we have $y_j\ll 1$ and $1-y_j\approx 1$. Then, by equations (\ref{sum_id}) and (\ref{w_cal}), we get
\begin{eqnarray}
  \langle w\rangle_R \approx\frac{15}{4\pi^4}e^{-s/2}\sum_{k=0}^{N-1}\int_{x_{k+1}}^{x_k}{\cal C}_{J_k}^{-1}\left[1-\exp\left(-\sum_{i=J_k}^N\tau_i\right)\right]f_{dB}(x)dx \;. \label{w_mean_01}
\end{eqnarray}

Since $x_1=h\nu_1/kT\ll 1$, the dominant contribution to the integral of $\langle w\rangle_R$ in equation (\ref{w_mean_01}) comes from the frequency segment $[x_1, \infty]$, i.e., the segment of $k=0$. Then, from equation (\ref{w_mean_01}) we get
\begin{eqnarray}
  \langle w\rangle_R \approx\frac{15}{4\pi^4}e^{-s/2}\int_{x_1}^\infty{\cal C}_1^{-1}\left[1-\exp\left(-\sum_{i=1}^N\tau_i\right)\right]f_{dB}(x)dx \;. \label{w_mean_02}
\end{eqnarray}
By equations (\ref{y_tau_j}) and (\ref{h_kap_j}), we have
\begin{eqnarray}
  \sum_{i=1}^N\tau_i=\alpha_{N,1}\left(\frac{h\nu_1}{kT}\right)^2(1-e^{-x})x^{-2} \;, \hspace{1cm}
  \alpha_{N,1}\equiv \frac{s}{\kappa_\es}\frac{\pi e^2}{m_e c}\sum_{j=1}^N\frac{n_{1,j}}{\rho_j}\frac{f_{12,j}}{\nu_j}\left(\frac{\nu_j}{\nu_1}\right)^2 \;. \label{alp_j_x1}
\end{eqnarray}

As $T\rightarrow\infty$, we have $\sum_{i=1}^N\tau_i\rightarrow 0$ since it is $\propto T^{-2}$. Hence, we can consider the limit of $\sum_{i=1}^N\tau_i\ll 1$. Then, we have ${\cal C}_1\approx 1-e^{-s/2}$, and $1-\exp\left(-\sum_{i=1}^N\tau_i\right)\approx\sum_{i=1}^N\tau_i$. In this limit, we get
\begin{eqnarray}
  \langle w\rangle_R \approx\frac{15}{4\pi^4}\frac{\alpha_{N,1}}{e^{s/2}-1}\left(\frac{h\nu_1}{kT}\right)^2\int_{x_1}^\infty(1-e^{-x})x^{-2}f_{dB}(x)dx \;. 
\end{eqnarray}

Since
\begin{eqnarray}
  \int_{x_1}^\infty(1-e^{-x})x^{-2}f_{dB}(x)dx\approx\int_0^{\infty}\frac{x^2}{e^x-1}dx=2\zeta(3) \;,
\end{eqnarray}
we get
\begin{eqnarray}
  \langle w\rangle_R \approx \frac{15\zeta(3)}{2\pi^4}\frac{\alpha_{N,1}}{e^{s/2}-1}\left(\frac{h\nu_1}{kT}\right)^2 \;, \hspace{1cm} kT\rightarrow\infty \;.  \label{wR_Ti1}
\end{eqnarray}

Equation~(\ref{wR_Ti1}) is valid in the limit of $kT\gg h\nu_1$ and $\alpha_{N,1}(h\nu_1/kT)^2\ll 1$. The second condition, i.e., $kT\gg h\nu_1\alpha_{N,1}^{1/2}$, is equivalent to the requirement that the total line optical depth for a photon of frequency $\sim kT/h$ is $\ll 1$.

Now, let us consider another limiting case: $kT\gg h\nu_1$ but $kT\ll h\nu_1\alpha_{N,1}^{1/2}$ (i.e., $kT$ is large but not $\rightarrow\infty$), which is possible only if $\alpha_{N,1}\gg 1$. Then, we have $\sum_{i=1}^N\tau_i\gg 1$, $1-\exp\left(-\sum_{i=1}^N\tau_i\right)\approx 1$, and ${\cal C}_1\approx 1$. By equation (\ref{w_mean_02}) we then get
\begin{eqnarray}
  \langle w\rangle_R \approx\frac{15}{4\pi^4}e^{-s/2}\int_{x_1}^\infty f_{dB}(x)dx\approx e^{-s/2} \;.  \label{wR_Ti2}
\end{eqnarray}
Hence, if $h\nu_1\ll kT\ll h\nu_1\alpha_{N,1}^{1/2}$, we have $\langle w\rangle_R\approx e^{-s/2}$ which is independent of $T$. Of course, if $\alpha_{N,1}\la 1$, the temperature region with a flat $\langle w\rangle_R$ does not exist.

Therefore, when $T\rightarrow 0$, in the integral of $\langle w\rangle_R$ only the last segment $[x_N,x_{N-1}]$ makes the dominant contribution, and we have that $\langle w\rangle_R$ approaches zero by $\propto (h\nu_N/kT)^4\exp(-h\nu_N/kT)$. When $T\rightarrow\infty$, only the first segment $[x_1,x_0=\infty]$ makes the dominant contribution, and we have that $\langle w\rangle_R$ approaches zero by $\propto (h\nu_1/kT)^{-2}$. The maximum of $\langle w\rangle_R$ is attained when the total contribution from all the segments between $x=x_N$ and $x=x_1$ becomes comparable to the contribution from the segment $[x_1,\infty]$.

\section{Numerical Results}
\label{num_result}

In this section we test the formulae for calculation of the line expansion opacity with the data of atomic lines extracted from the Atomic Spectral Line Database from CD-ROM 23 of R. L. Kurucz.\footnote{https://www.cfa.harvard.edu/amp/ampdata/kurucz23/sekur.html} We calculate the monochromatic line expansion opacity with equations (\ref{kap_exp1a})--(\ref{cal_CJ}), the Rosseland mean of the enhancement factor  with equation (\ref{w_cal}), and then the Rosseland mean of the line expansion opacity with equation (\ref{ros_w}). Notice that the line expansion opacity in equation (\ref{kap_exp1a}) and its Rosseland mean in equation~(\ref{ros_w}) include the contribution of electron scattering. The pure line expansion opacity is related to the $w$-factor by
\begin{eqnarray}
  \kappa_{\lex}\equiv\kappa_{\expp}-\kappa_\es=\kappa_\es\frac{w}{1-w} \;, \label{kap_w2}
\end{eqnarray}
where we use the subscript ``lex'' to denote ``line expansion''. Similarly, we define the Rosseland mean of the pure line expansion opacity by
\begin{eqnarray}
  \kappa_{\lex,R}\equiv \kappa_{\exp,R}-\kappa_\es=\kappa_\es\frac{\langle w\rangle_R}{1-\langle w\rangle_R} \;. \label{kap_wr}
\end{eqnarray}

We use the atomic spectral lines of neodymium (Nd) and iron (Fe) to test the formulae. Neodymium is a typical lanthanide element, whose atomic lines have been claimed to play a dominant role in increasing the opacity in the ejecta of a neutron star merger \citep{kas13,tan13}. Iron and elements near iron in the periodic table are common in the ejecta of supernovae, and are important in determining the opacity, the luminosity, and the spectral characteristics of supernovae. From the database we extracted the line data of Nd~II, Fe~IV, and Fe~III. Nd~II is the singly ionized state of neodymium, also the only ionized state of neodymium with atomic lines contained in the database. The data of Nd~II contain in total 1002 spectral lines covering the wavelength range of 299.3--883.9\,\nm. Fe~IV is the triply ionized iron, which has in total 7897 spectral lines with wavelength in the range of $35.6$--$8293.5\,\nm$. Fe~III is the doubly ionized iron, which has in total 23,059 spectral lines with wavelength in the range of $45.5$--$96,146\,\nm$.

\begin{figure}[ht!]
\centering{\includegraphics[angle=0,scale=0.8]{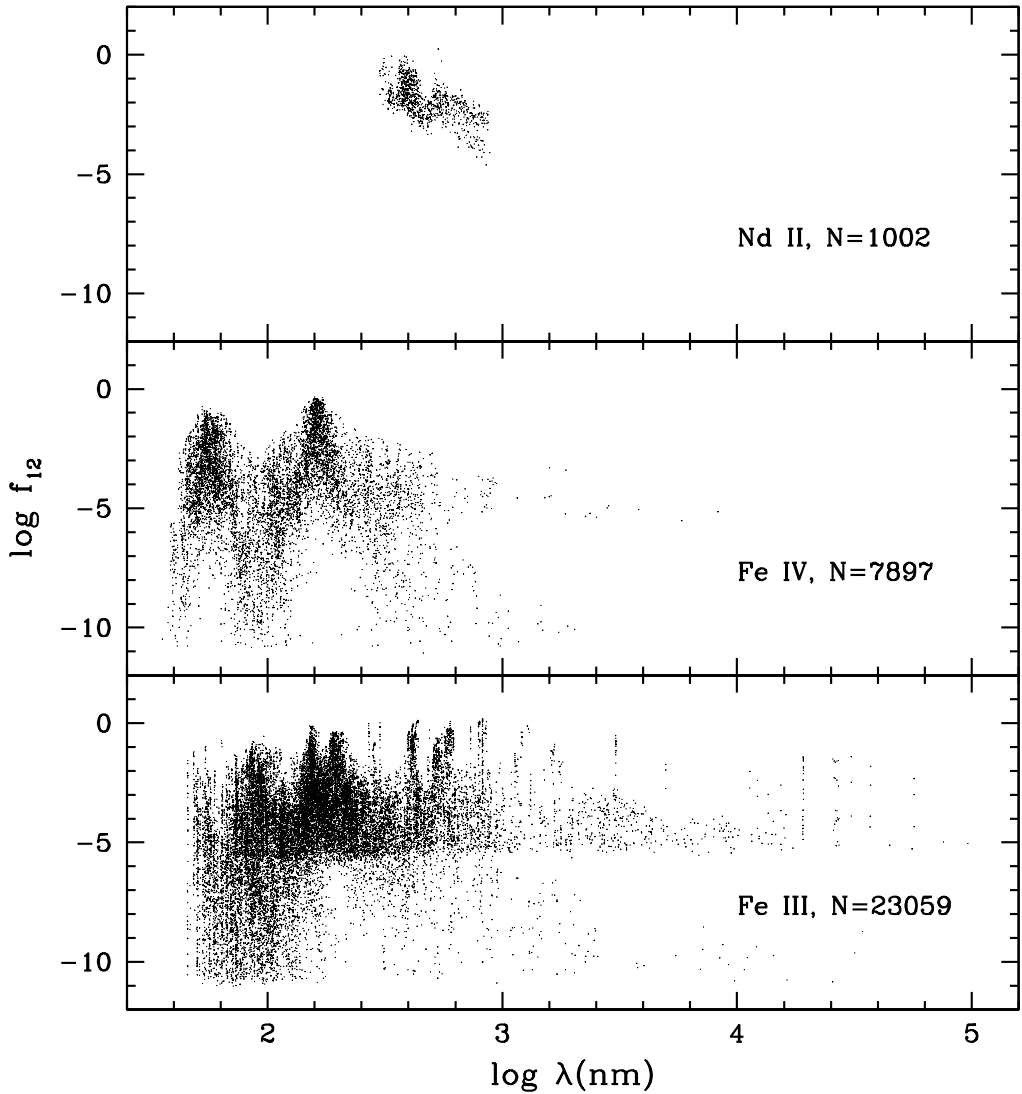}}
\caption{The emission oscillator strength versus the wavelength for the spectral lines of Nd~II, Fe~IV, and Fe~III, where $N$ is the total number of lines for each element.
}
\label{f_lam}
\end{figure}

In Fig.~\ref{f_lam} we plot the emission oscillator strength versus the wavelength for the spectral lines of Nd~II, Fe~IV, and Fe~III. Among the three elements, Nd~II has the smallest number (1002) of lines, which is just enough for testing the formulae that we have derived. The wavelength of the lines is distributed in a narrow range, with the emission oscillator strength $>10^{-5}$ for all the lines. Fe~IV has a larger number (7897) of lines than Nd~II, with the wavelength distributed in a wider range. Fe~III has the largest number (23,059) of lines among the three, with the wavelength distributed in an even wider range than Fe~IV. Calculation with an increasing number of spectral lines of different elements allows us to check the variation of the line expansion opacity with respect to element species and line numbers, and test the efficiency of the computer code used to evaluate the line expansion opacity and its Rosseland mean. Compared to Nd~II, Fe~IV and Fe~III have many weak lines with the emission oscillation strength $<10^{-5}$ down to $10^{-12}$. As we have discussed in Section \ref{discuss1}, those many weak lines can have important contribution to the opacity hence cannot be ignored.

\begin{figure}[ht!]
\centering{\includegraphics[angle=0,scale=0.7]{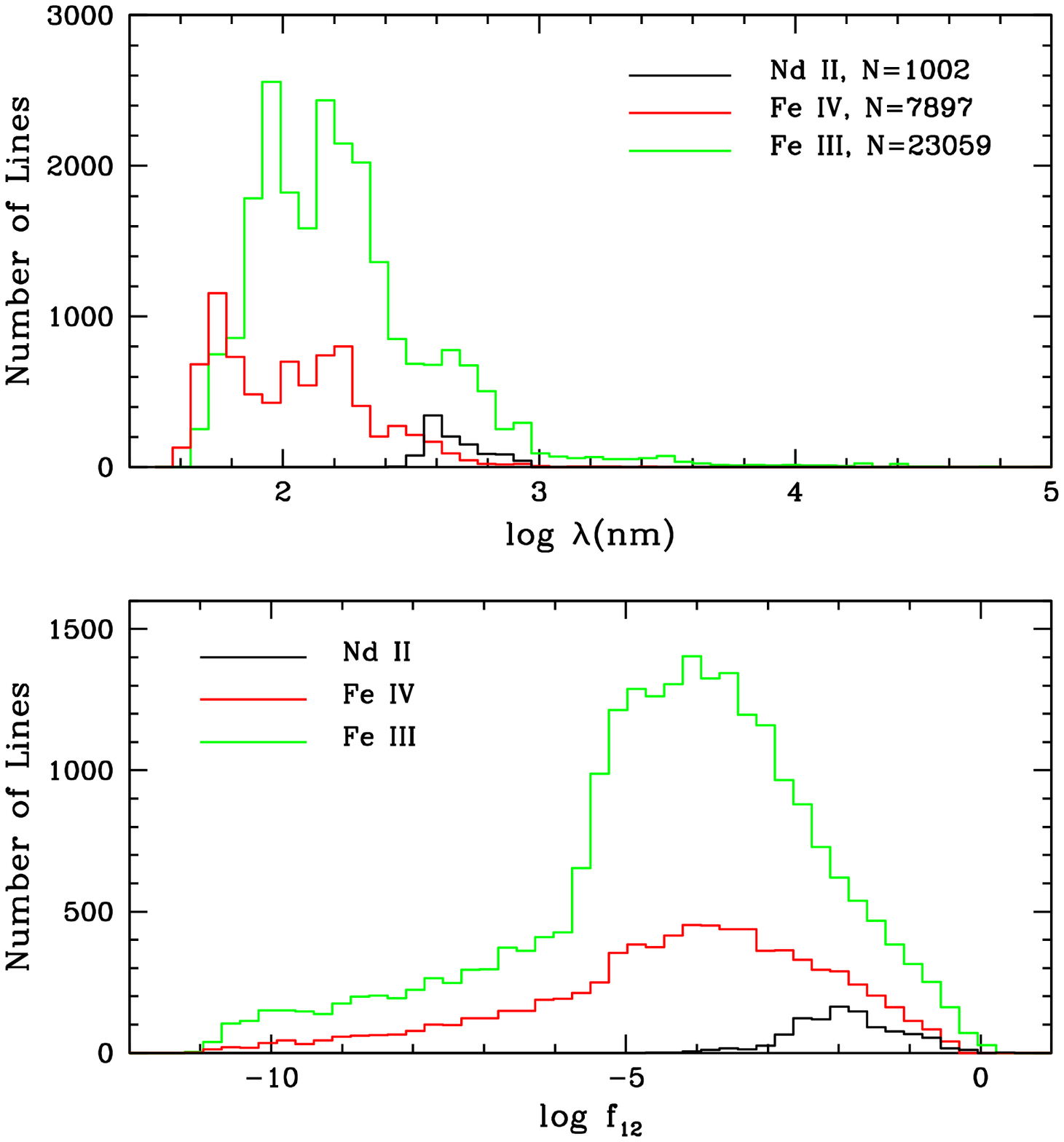}}
\caption{Histogram distributions of the number of atomic spectral lines over the wavelength (upper panel) and the emission oscillator strength (lower panel) for Nd~II, Fe~IV, and Fe~III, where $N$ in the upper panel is the total number of lines for each atomic element. 
}
\label{hist}
\end{figure}

Distributions of the number of atomic spectral lines over the wavelength and the emission oscillator strength for each of the three elements are shown in Fig.~\ref{hist}. From the figure we see that, relative to that of Fe~IV and Fe~III, the wavelength of the lines of Nd~II is distributed in a significantly narrower range, peaking at a significantly longer wavelength. This difference has an important effect on the line expansion opacity, which will be manifested in the numerical results presented below. From Fig.~\ref{hist} we also see that for Fe~IV and Fe~III the emission oscillator strength is peaked around $10^{-4}$, with a long tail in the range of $10^{-12}-10^{-5}$. While for Nd~II, the emission oscillator strength is peaked around $10^{-2}$, and all are beyond $10^{-5}$. This fact confirms an impression that we have obtained from Fig.~\ref{f_lam}: Nd~II contains dominantly strong lines, while Fe~IV and Fe~III contain both strong lines and a lot of weak lines.

The results for the line expansion opacity and its Rosseland mean evaluated with the spectral data of Nd~II, Fe~IV, and Fe~III are presented below. We emphasize that the purpose of the numerical calculations presented here is only for testing the formulae that we have derived, not for modeling the opacity in a realistic situation of supernovae and neutron star mergers. Therefore we do not even care if the data are complete and accurate. The chemical compositions of a supernova and a neutron star merger are very complicated, which cannot be modeled by the few atomic elements outlined above. For instance, even though people have predicted that neodymium can dominate the line expansion opacity in a neutron star merger, here we calculate only the line expansion opacity arising from its first ionized state (Nd~II). Higher ionization sates of neodymium are not included in our calculations due to the unavailability of data in the database mentioned above. For the purpose of testing the formulae for the line expansion opacity and its Rosseland mean, the calculations presented here are enough.

\subsection{The result for Nd~II}
\label{Nd_II}

\begin{figure}[ht!]
\centering{\includegraphics[angle=0,scale=0.7]{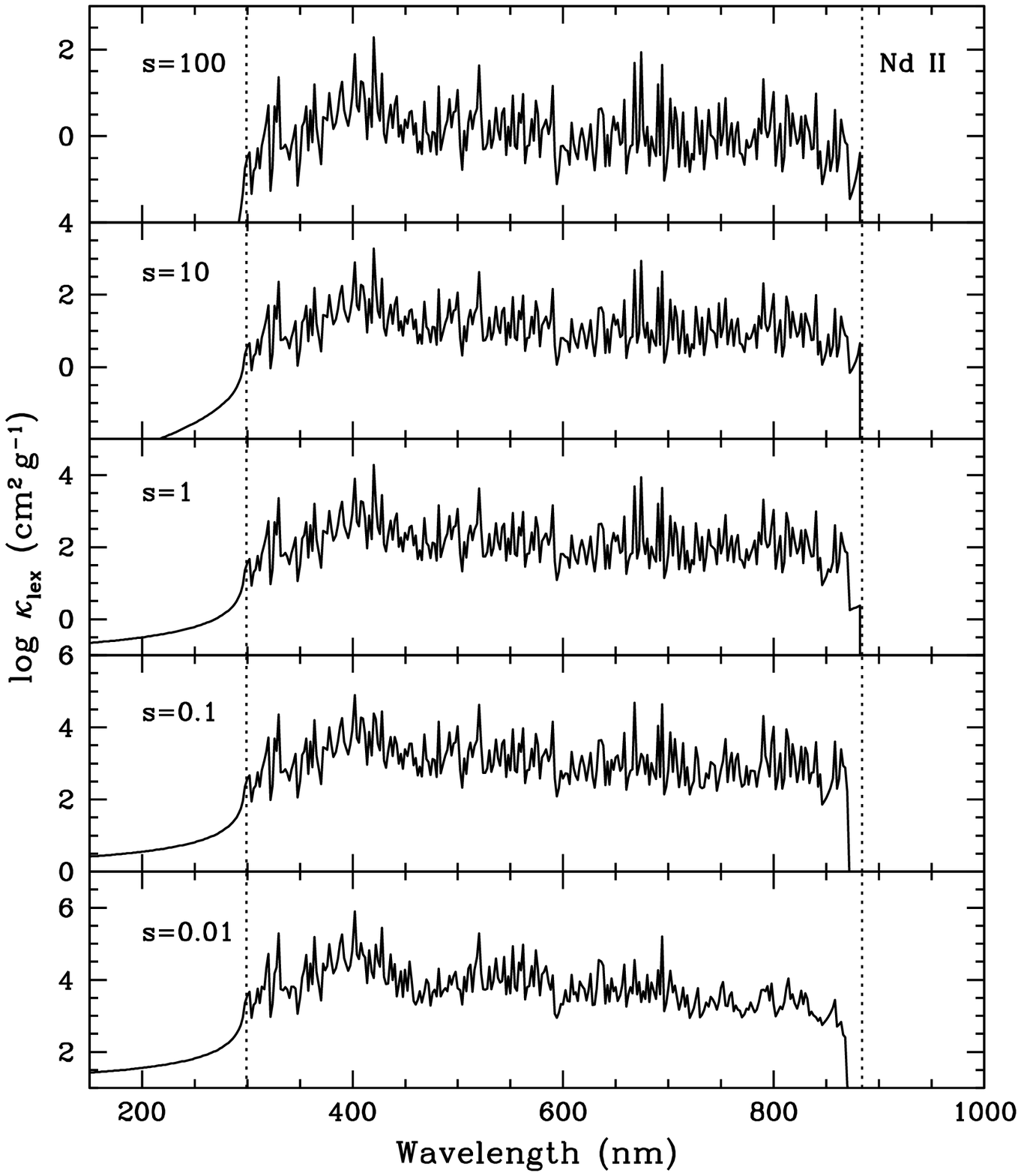}}
\caption{The monochromatic line expansion opacity of Nd~II for various values of the expansion parameter $s$. The two vertical dotted lines mark the minimum and maximum wavelengths of the spectral lines. The temperature of the medium is taken to be $6000\,\K$. The opacity of electron scattering is $\kappa_\es= 0.1\,\cm^2\,\g^{-1}$. The mass abundance of Nd~II in the spherical matter is $0.01$.
}
\label{lopc1}
\end{figure}

The monochromatic line expansion opacity generated by the spectral lines of Nd~II in a uniformly expanding sphere is calculated and shown in Fig.~\ref{lopc1} for various values of the expansion parameter $s$. We assume that the uniform spherical matter has a temperature of $6000\,\K$ at some given moment, the opacity of electron scattering is $\kappa_\es=0.1\,\cm^2\,\g^{-1}$, and Nd~II occupies $1\%$ of the total mass in the sphere.

In the line expansion opacity presented in Fig.~\ref{lopc1} the contribution from the electron scattering has been removed. That is, the pure line expansion opacity defined in equation (\ref{kap_w2}) is presented. The minimum of the wavelength is $\lambda_{\min}=c/\nu_1$, and the maximum is $\lambda_{\max}=c/\nu_N$. For $\lambda>\lambda_{\max}$, we have $\nu<\nu_N$, the frequency of a photon in this range will be redshifted away from the spectral lines and will not interact with any of them. Hence, for $\lambda>\lambda_{\max}$ the line expansion opacity is zero. For $\lambda<\lambda_{\min}$, the photon has a frequency $\nu>\nu_1$, which will interact will all the spectral lines as the matter expands because of the Doppler effect. Hence, for $\lambda<\lambda_{\min}$ the line expansion opacity has a nonzero tail, as shown in Fig.~\ref{lopc1}. The line expansion opacity decreases with decreasing $\lambda$ for $\lambda<\lambda_{\min}$, caused by the factor $\exp[-(s/2)(1-\nu_i^2/\nu^2)]$ in the equation for the line expansion opacity (eq.~\ref{kap_exp1b}). Obviously, the larger the value of $s$, the faster the $\kappa_{\lex}$ decreases with decreasing $\lambda$.

\begin{figure}[ht!]
\centering{\includegraphics[angle=0,scale=0.7]{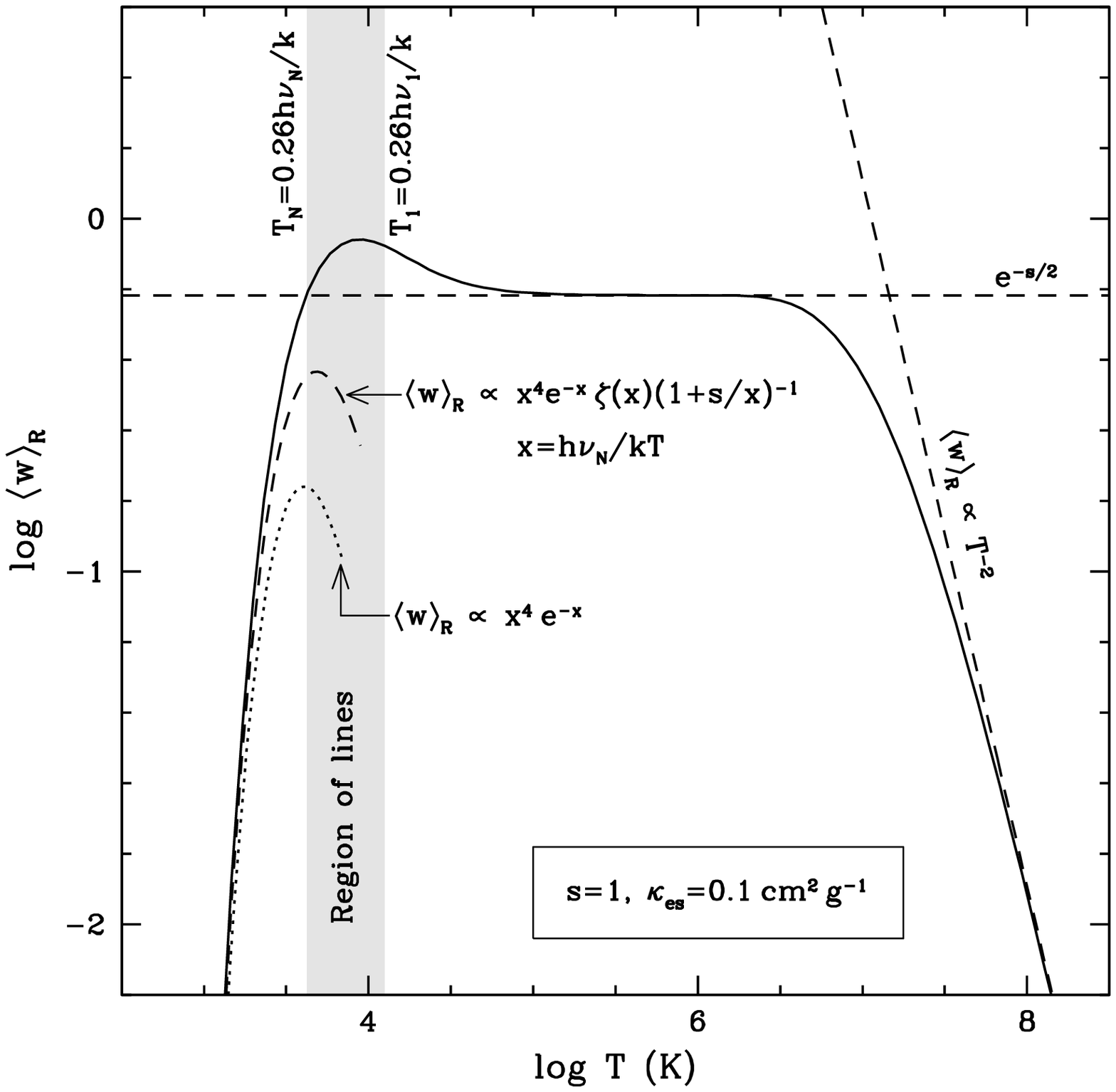}}
\caption{Rosseland mean of the enhancement factor, $\langle w\rangle_R$, as a function of the temperature $T$ for the case of $s=1$ and Nd~II. The asymptotic solutions for $T\rightarrow 0$ and $T\rightarrow \infty$, and the approximate constant solution when $h\nu_1/k\ll T\ll\alpha_{N,1}^{1/2}h\nu_1/k$, are shown with dashed and dotted curves (see Secs.~\ref{asym1} and \ref{asym2}, respectively). The region containing the spectral lines are shaded, where the temperature of a line is defined by $T_i=x_m^{-1}h\nu_i/k=0.261h\nu_i/k$, $1\le i\le N$.
}
\label{wr1}
\end{figure}

To see how the Rosseland mean of the line expansion opacity varies with temperature, in Fig.~\ref{wr1} we plot the Rosseland mean of the enhancement factor, $\langle w\rangle_R$, as a function of temperature for the case of $s=1$. The asymptotic solutions as $T\rightarrow 0$ and $T\rightarrow\infty$ derived in Sections \ref{asym1} and \ref{asym2} are also shown. The shaded region marks the approximate location of the spectral lines in the temperature space, with the left boundary defined by $T_N=x_m^{-1}h\nu_N/k=0.261h\nu_N/k$ and the right boundary defined by $T_1=0.261h\nu_1/k$. We see that, the numerical results approach the asymptotic solutions correctly. According to the discussion in Sections \ref{asym1} and \ref{asym2}, as $T\rightarrow0$ we have $\langle w\rangle_R\propto x^4e^{-x}$, where $x=h\nu_N/kT$. As $T\rightarrow\infty$, we have $\langle w\rangle_R\propto(kT)^{-2}$. While in the interval from $T\sim h\nu_1/k$ to $T\sim \alpha_{N,1}^{1/2}h\nu_1/k$, $\langle w\rangle_R$ has a roughly constant value $\approx e^{-s/2}$ when $\alpha_{N,1}\gg 1$. When $s$ is very large, $e^{-s/2}$ is sufficiently small so that the flat part in the $\langle w\rangle_R$-$T$ curve will not be visible. When $s$ is small, we may have $\alpha_{N,1}<1$ and in this case the flat part will also be absent (see Fig.~\ref{mopc1} below). 

\begin{figure}[ht!]
\centering{\includegraphics[angle=0,scale=0.7]{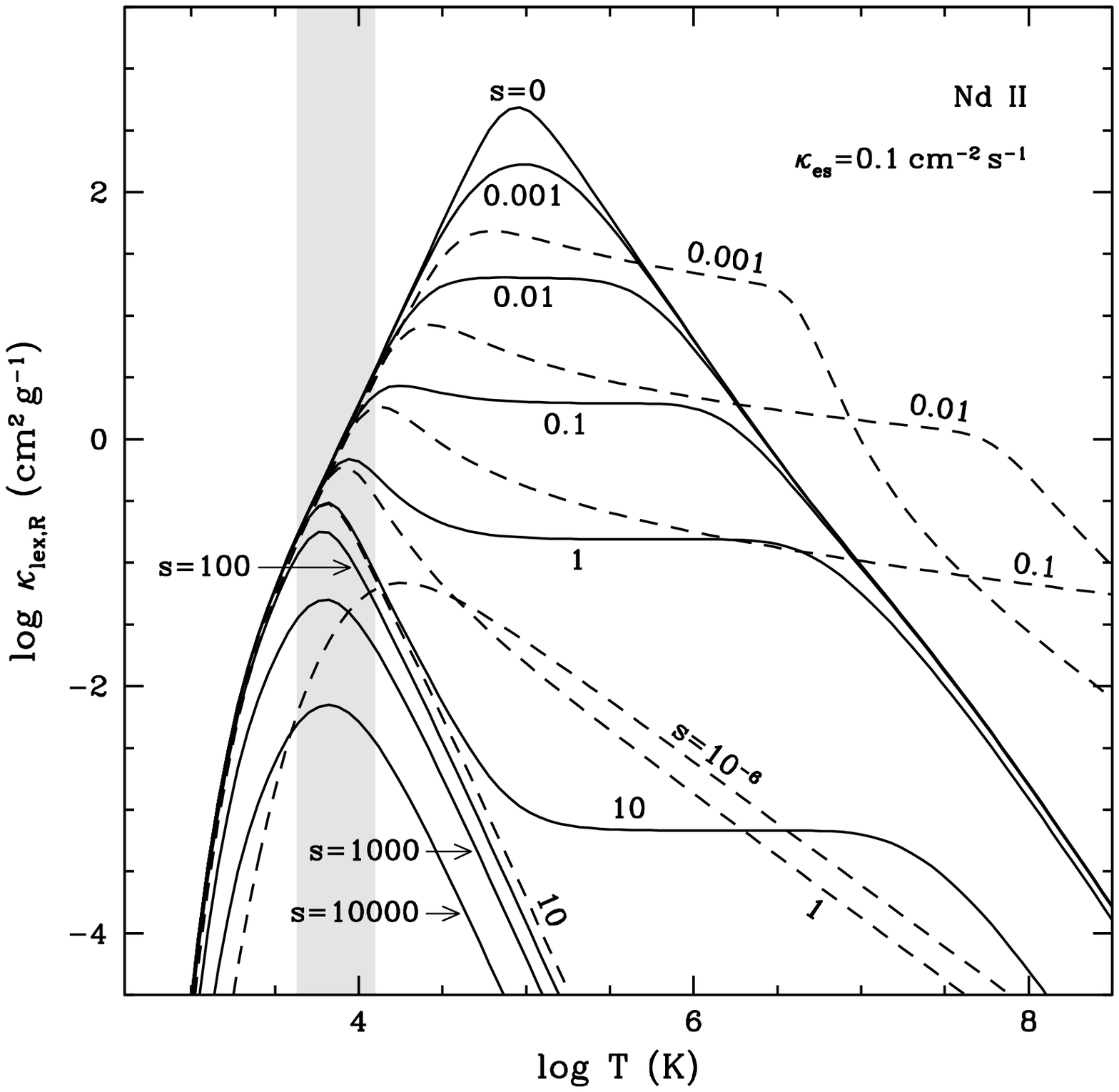}}
\caption{Rosseland mean of the line expansion opacity as a function of temperature for various values of the expansion parameter $s$. The results calculated for Nd~II with the formulae derived in this work, i.e., equations (\ref{w_cal}) and (\ref{kap_wr}), are displayed with solid curves. The expansion parameter $s$ is taken to be equal to 10,000, 1000, 100, 10, 1, 0.1, 0.01, 0.001, and 0, as denoted on the curves. The opacity of electron scattering is assumed to be $\kappa_\es=0.1\,\cm^{-2}\,\g^{-1}$. The mass abundance of Nd~II in the spherical matter is $0.01$. For comparison, the results calculated with the formulae derived by \citet{kar77} are shown with dashed curves, with $s=10$, 1, 0.1, 0.01, 0.001, and $10^{-6}$. For $s=100$, 1000, and 10,000, the results given by the formulae of \citet{kar77} are almost indistinguishable from that calculated with our formulae for the temperature range shown in the figure so they are not shown. The gray shaded region shows the location of spectral lines in the space of temperature, with the temperature of a line of frequency $\nu_l$ being defined by $T_l=0.261h\nu_l/k$ (see the text).
}
\label{mopc1}
\end{figure}

The Rosseland mean of the line expansion opacity for Nd~II calculated with our formulae are shown in Fig.~\ref{mopc1} for various values of the expansion parameter $s$ ranging from 10,000 to 0 (solid curves). For comparison, the results calculated with the formulae of {\klcs} are also shown (dashed curves). For $s\ga 100$, the results calculated with the formulae of {\klcs} are very close to that calculated with our formulae, the two are almost indistinguishable in the figure. However, for $s<100$, the results given by our formulae are clearly distinguishable from that given by the formulae of {\klcs}. After the peak, the Rosseland mean of the line expansion opacity calculated with our formulae decays with increasing temperature at a rate slower than that calculated with the formulae of {\klcs} when the flat region is present. The line expansion opacity calculated with our formulae increases monotonically with decreasing $s$ and reaches a finite limit at $s=0$, while the line expansion opacity calculated with the formulae of {\klcs} increases with decreasing $s$ for $s\la 0.001$ then decreases as $s$ decreases further. For instance, when $s=10^{-6}$, our formulae lead to a line expansion opacity between that of $s=0.001$ and that of $s=0$, but the formulae of {\klcs} lead to a line expansion opacity much smaller than that of $s=0.001$. As $s\rightarrow 0$, the formulae of {\klcs} lead to a zero line expansion opacity.

\begin{figure}[ht!]
\centering{\includegraphics[angle=0,scale=0.7]{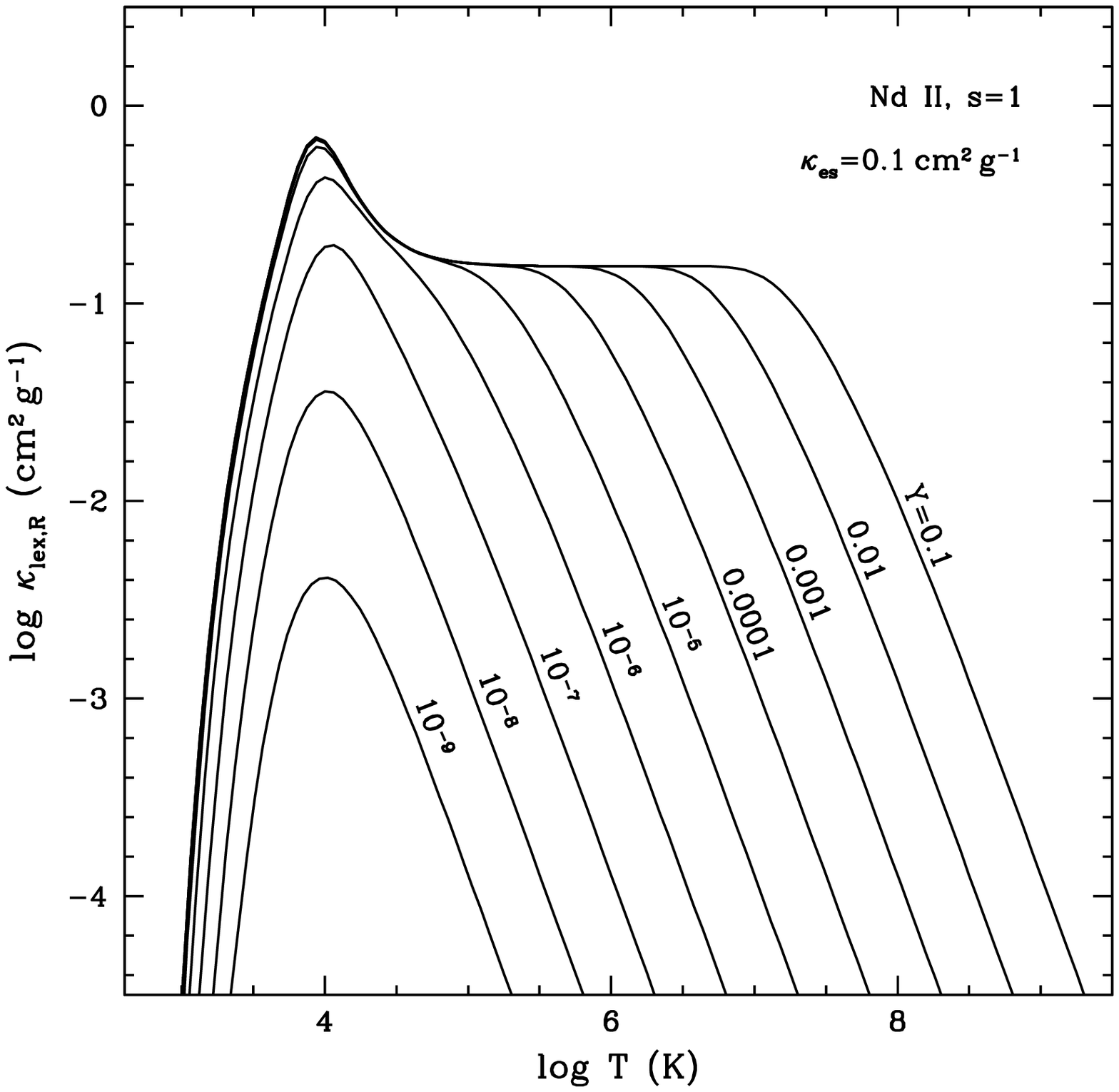}}
\caption{Variation of the Rosseland mean of the line expansion opacity, $\kappa_{\lex,R}$, with respect to the mass abundance of Nd~II for the case of $s=1$. The mass abundance $Y$ is taken to be equal to 0.1, 0.01, 0.001, 0.0001, $10^{-5}$, $10^{-6}$, $10^{-7}$, $10^{-8}$, and $10^{-9}$, as denoted on the curves. The opacity of electron scattering is assumed to be $\kappa_\es=0.1\,\cm^{-2}\,\g^{-1}$. The horizontal axis is the temperature of the medium.
}
\label{mopc1_fr}
\end{figure}

Variation of the Rosseland mean of the line expansion opacity with respect to the abundance of Nd~II is shown in Fig.~\ref{mopc1_fr}. The trend agrees with theoretical expectations. By equation (\ref{wR_Ti1}), as $T\rightarrow\infty$ we have $\langle w\rangle_R\propto\alpha_{N,1}\propto n_1$, hence $\kappa_{\lex,R}\propto Y$ where $Y$ is the mass abundance of the atomic element. According to equation (\ref{wR_Ti2}), the flat part in the $\kappa_{\lex,R}$-$T$ curve does not depend on the element abundance. According to equation (\ref{w_T_lim1a}), as $T\rightarrow 0$, we have $\langle w\rangle_R\propto\left(1-e^{-\tau_N}\right)/\left(1-e^{-\tau_N-s/2}\right)$. When $\tau_N$ is large, we have $\left(1-e^{-\tau_N}\right)/\left(1-e^{-\tau_N-s/2}\right)\approx 1$, $\langle w\rangle_R$ and $\kappa_{\lex,R}$ do not vary when the element abundance changes. When $\tau_N$ is small, we have $\left(1-e^{-\tau_N}\right)/\left(1-e^{-\tau_N-s/2}\right)\approx \tau_N\propto n_1$ and then $\langle w\rangle_R$ and $\kappa_{\lex,R}\propto Y$. The above analyses agree with the numerical results shown in Fig.~\ref{mopc1_fr}.

Since $\tau_j\propto n_{1,j}\propto Y$, as $Y\rightarrow 0$ we have $\tau_j\rightarrow 0$. By equation (\ref{w_cal}), as $\tau_j\rightarrow 0$ we have
\begin{eqnarray}
  \langle w\rangle_R \approx \frac{15}{4\pi^4\left(e^{s/2}-1\right)}\sum_{k=0}^{N-1}\sum_{j=J_k}^N\int_{x_{k+1}}^{x_k}\tau_j(x)e^{sy_j(x)/2}f_{dB}(x)dx \;. \label{w__sY}
\end{eqnarray}
Hence, as $Y\rightarrow 0$ we have
\begin{eqnarray}
  \kappa_{\lex,R}\approx \kappa_{\es}\langle w\rangle_R\propto Y \;, \label{kap_sY}
\end{eqnarray}
which agrees with the numerical result in Fig.~\ref{mopc1_fr}.

\subsection{The result for Fe~IV and Fe~III}
\label{Fe}

\begin{figure}[ht!]
\centering{\includegraphics[angle=0,scale=0.7]{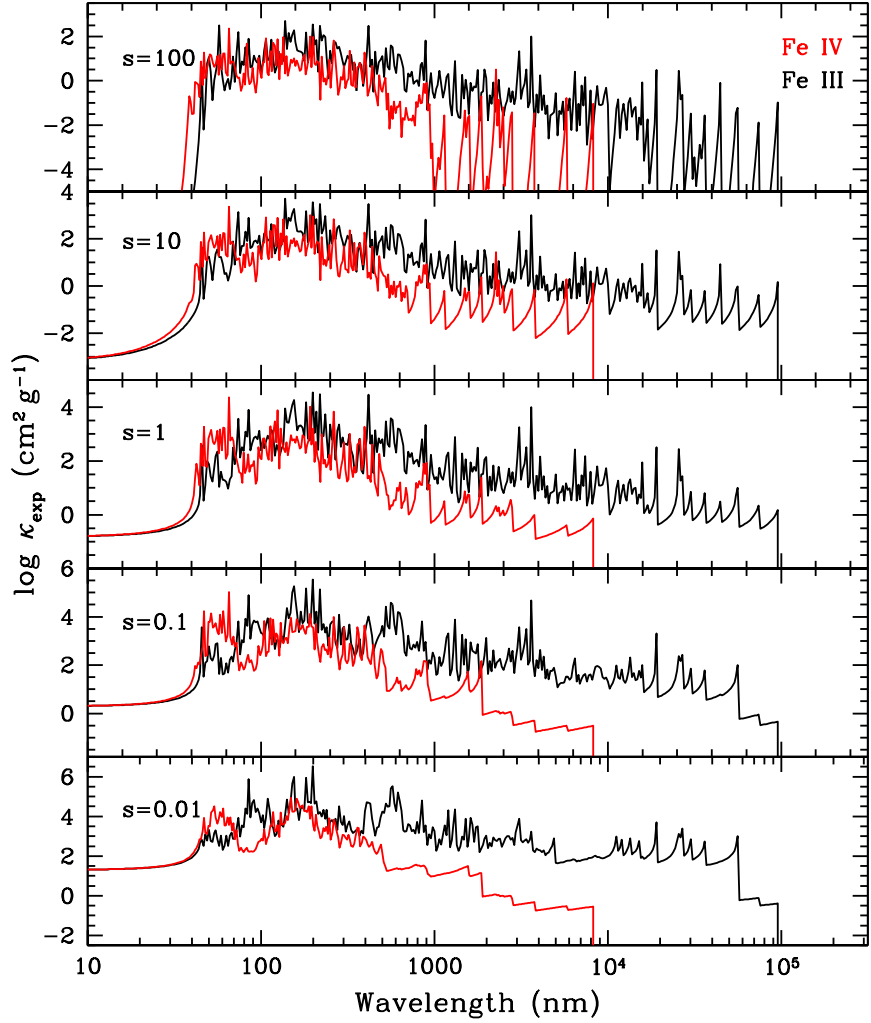}}
\caption{Similar to Fig.~\ref{lopc1} but for Fe~IV (red) and Fe~III (black). Since the range of the wavelength spans more than two orders of magnitude, in this figure we show the wavelength in logarithm scales.
}
\label{lopc2}
\end{figure}

\begin{figure}[ht!]
\centering{\includegraphics[angle=0,scale=0.7]{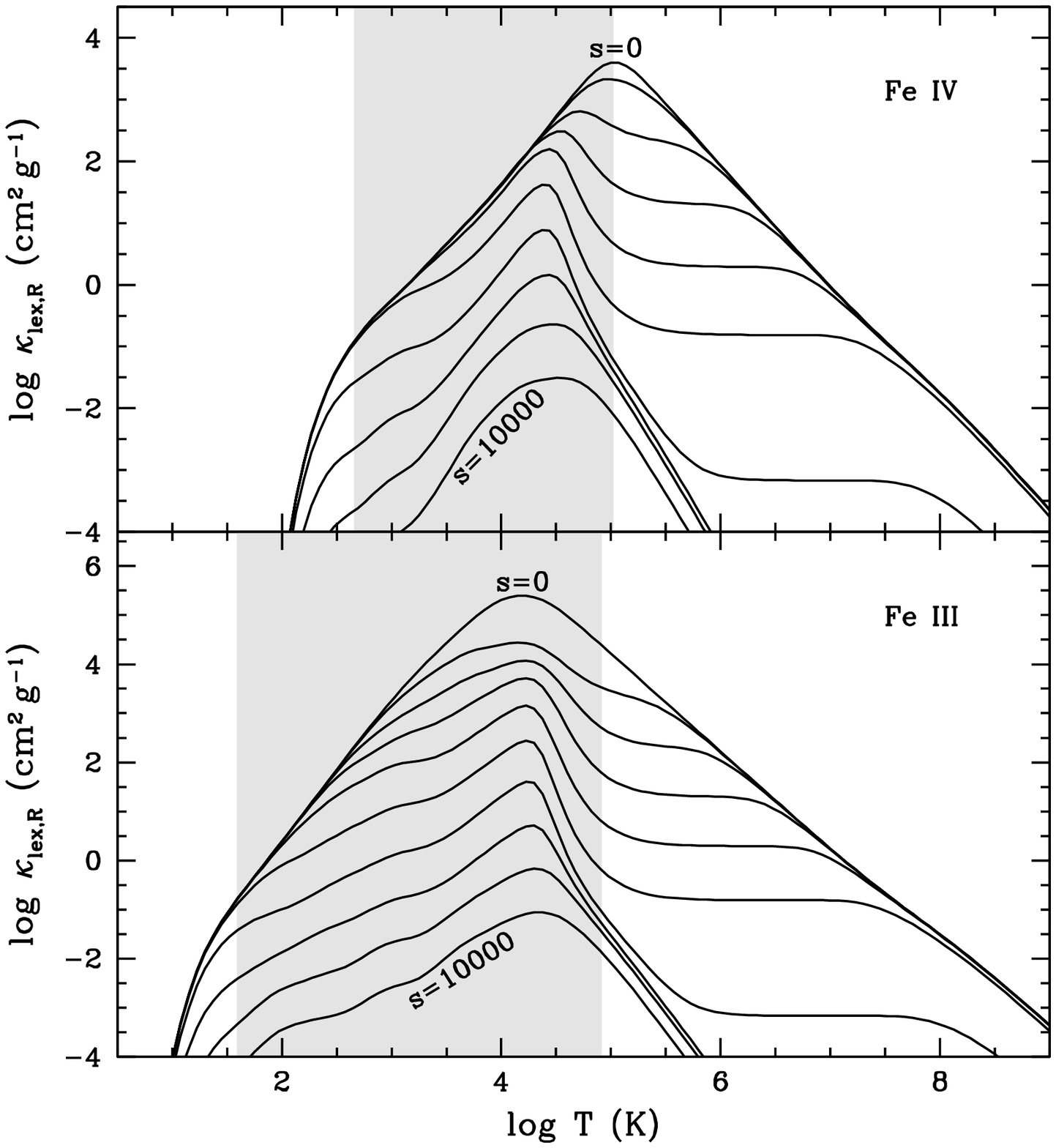}}
\caption{Similar to Fig.~\ref{mopc1} but for Fe~IV (upper panel) and Fe~III (lower panel). Only the results calculated with the formulae derived in this paper are shown. Top to bottom in each panel: $s=0$, 0.0001, 0.001, 0.01, 0.1, 1, 10, 100, 1000, and 10,000.
}
\label{mopc2}
\end{figure}

We have calculated the line expansion opacity and its Rosseland mean for elements with more complex spectral lines, Fe~IV and Fe~III, with our formulae. The results are shown in Figs.~\ref{lopc2} and \ref{mopc2}. As in the case of Nd II, we assume that the opacity of electron scattering is $\kappa_\es=0.1\,\cm^2\,\g^{-1}$, and the mass of Fe~IV and Fe~III is $1\%$ of the total mass of the sphere, respectively. In Fig.~\ref{lopc2} it is further assumed that the matter has a temperature of $6000\,\K$.

Overall, the results for Fe~IV and Fe~III are very similar to that for Nd~II, with some subtle differences. For the monochromatic line expansion opacity, as we have explained for Nd~II, the line expansion opacity vanishes for $\nu<\nu_N$ (i.e., for $\lambda>\lambda_N$) since a photon with such a low frequency will not interact with any spectral line in the line sample. The line expansion opacity does not vanish for $\nu>\nu_1$ (i.e., for $\lambda<\lambda_1$), since a photon with a frequency $\nu>\nu_1$ will have chances to interact with all spectral lines. Since Fe~IV and Fe~III have considerably more lines than Nd~II (7897 lines for Fe~IV and and 23,059 lines for Fe~III vs. 1002 lines for Nd~II), which span a much broader range in wavelength and frequency ($35.6$--$8293.5\,\nm$ for Fe~IV and $45.5$--$96,146\,\nm$ for Fe~III vs. $299.3$--$883.9\,\nm$ for Nd~II), under the same conditions it is expected that Fe~IV and Fe~III have a stronger effect on the line expansion opacity. This is confirmed by comparison of Fig.~\ref{lopc2} to Fig.~\ref{lopc1}, and comparison of the black lines to the red lines in Fig.~\ref{lopc2}. For instance, the line expansion opacity of Fe~IV clearly affects photons with a broader range of wavelength than that of Nd~II, and the line expansion opacity of Fe~III clearly affects photons with a broader range of wavelength than that of Fe~IV. However, when other parameters are the same, the amplitudes of the line expansion opacity for the three elements are similar.

The differences between the results for Fe~IV/Fe~III and Nd~II are more clearly seen by comparison of Fig.~\ref{mopc2} to Fig.~\ref{mopc1}, where the Rosseland mean of the line expansion opacity generated by Fe~IV/Fe~III and Nd~II are shown. Clearly, the Rosseland mean of the line expansion opacity of Fe~IV/Fe~III has a larger amplitude and affects a wider range of temperature than that of Nd~II. Fe~III has more spectral lines than Fe~IV, with the wavelength covering a broader range. Not surprisingly, Figs.~\ref{lopc2} and \ref{mopc2} show that Fe~III has a stronger effect on the line expansion opacity and its Rosseland mean than both Fe~IV and Nd~II. For instance, the Rosseland mean of the line expansion opacity of Fe~III has a larger amplitude and affects a wider range of temperature than that of Fe~IV.

\begin{figure}[ht!]
\centering{\includegraphics[angle=0,scale=0.7]{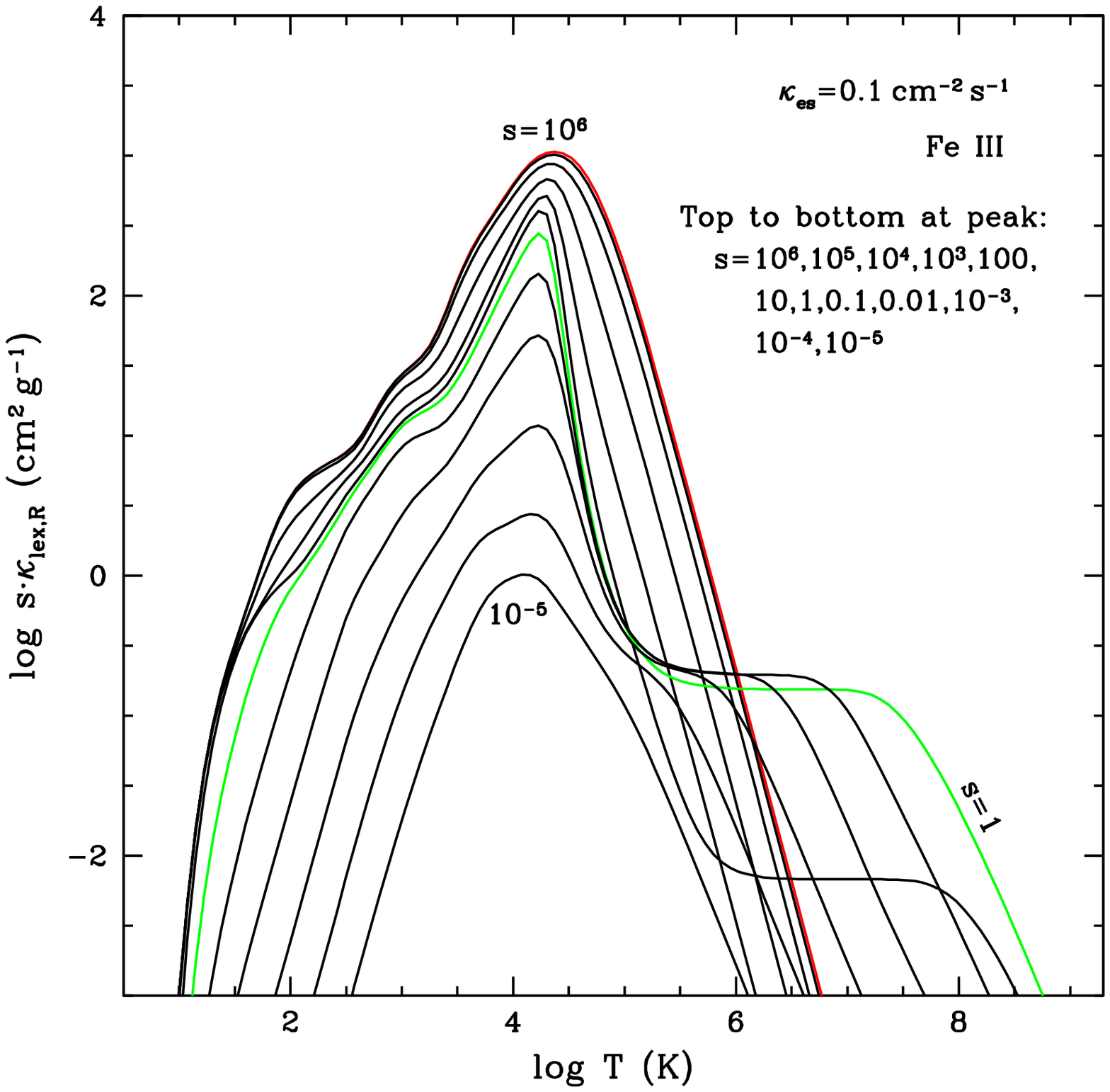}}
\caption{Plot of $s\kappa_{\lex,R}$ versus temperature for the case of Fe~III for various values of $s$. From top to bottom at the peak of the curve, $s=10^6$, $10^5$, $10^4$, $10^3$, $100$, 10, 1, 0.1, 0.01, $10^{-3}$, $10^{-4}$, and $10^{-5}$. When $s>10^6$, the curve is visually indistinguishable from that of $s=10^6$ in the figure due to the fact that $\kappa_{\lex,R}\propto s^{-1}$ when $s$ is large. The solutions for $s=10^{6}$ and $s=1$ are plotted with color curves so that they can be more easily identified.
}
\label{mopc3_s}
\end{figure}

\begin{figure}[ht!]
\centering{\includegraphics[angle=0,scale=0.7]{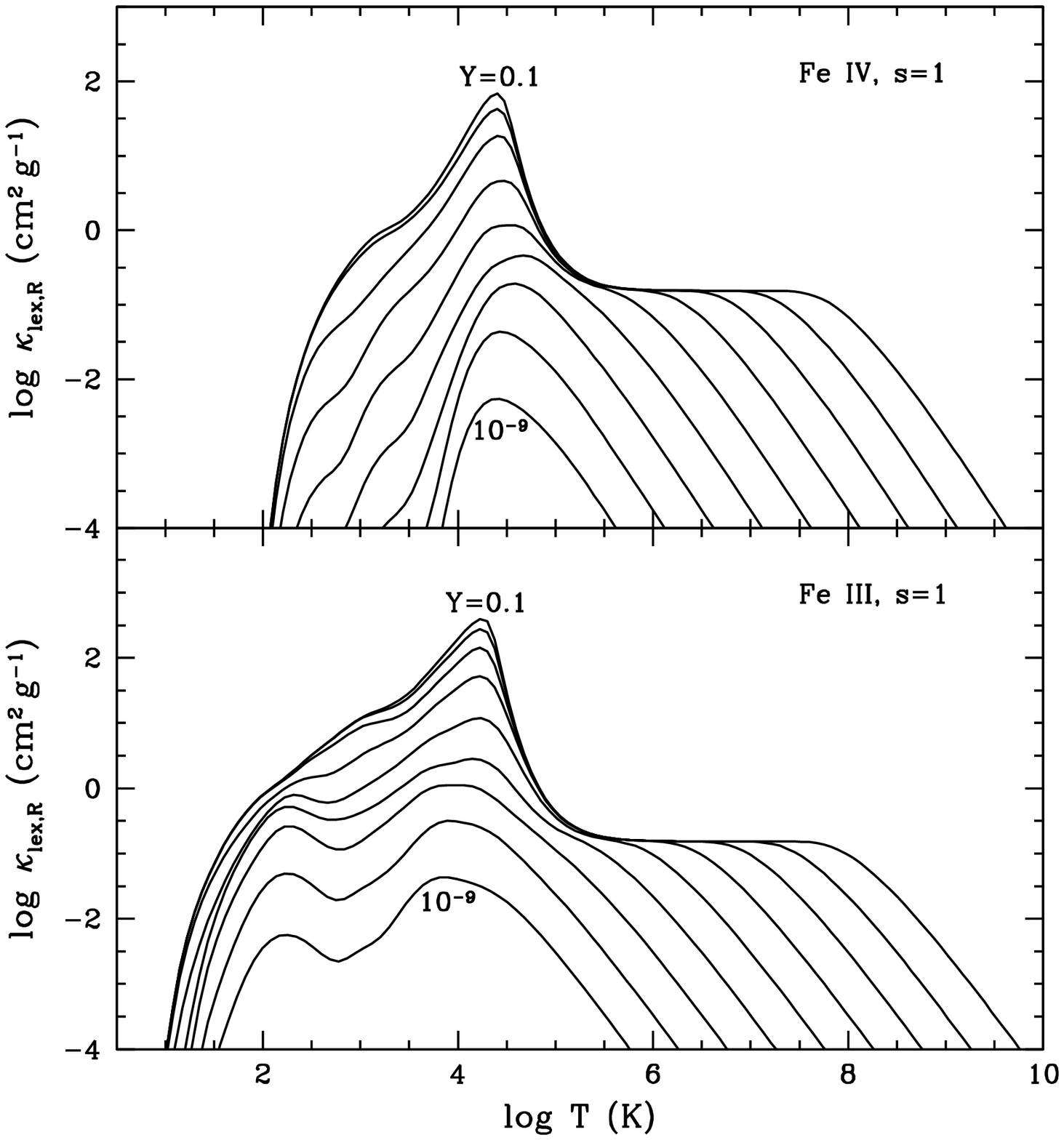}}
\caption{Similar to Fig.~\ref{mopc1_fr} but for the cases of Fe~IV (upper panel) and Fe~III (lower panel). From top to bottom in each panel: $Y=0.1$, 0.01, 0.001, 0.0001, $10^{-5}$, $10^{-6}$, $10^{-7}$, $10^{-8}$, and $10^{-9}$.
}
\label{mopc23_fr}
\end{figure}

Since $s\kappa_{\lex,R}/\kappa_\es=\kappa_{\lex,R}\rho c\eta$, $s\kappa_{\lex,R}/\kappa_\es$ is a measure of the optical depth associated with the line expansion opacity. In Fig.~\ref{mopc3_s} we plot $s\kappa_{\lex,R}$ versus temperature for the case of Fe~III. Although $\kappa_{\lex,R}$ can be very large when $s\ll 1$ as we have seen in Figs.~\ref{mopc1} and \ref{mopc2}, $s\kappa_{\lex,R}$ decreases with decreasing $s$. In fact, when $s\rightarrow 0$ we have $s\kappa_{\lex,R}\rightarrow 0$, since $\kappa_{\lex,R}\rightarrow {\cal O}(s^0)$ as $s\rightarrow 0$. As will be proved in Section \ref{s_inf} as $s\rightarrow\infty$ we have $\kappa_{\lex,R}\propto s^{-1}$, which is verified by Fig.~\ref{mopc3_s} since when $s>10^6$ the curve for $s\kappa_{\lex,R}$ becomes visually indistinguishable from that of $s=10^6$.

In Fig.~\ref{mopc23_fr} we show the variation of $\kappa_{\lex,R}(T)$ with respect to the mass abundance of Fe~IV (upper panel) and Fe~III (lower panel) for the case of $s=1$. Similar to the case of Nd~II (Fig.~\ref{mopc1_fr}), the results agree with the theoretical analyses presented in Section~\ref{Nd_II}, including the trend as $Y\rightarrow 0$ given by equation~(\ref{kap_sY}).

\subsection{Approximate solution at the peak}
\label{peak}

From Figs.~\ref{mopc1} and \ref{mopc2} we can see an interesting difference in the feature of the Rosseland mean of the line expansion opacity for different elements. For Fe~IV and Fe~III, the peak of $\kappa_{\lex,R}$ is always within the range of the spectral lines (the gray shaded region). However, for Nd~II, when $s<1$ the peak of $\kappa_{\lex,R}$ occurs at temperature that is outside the line region, i.e., at temperature $T_p>x_m^{-1}h\nu_1/k$. This difference in the behavior of the peak of $\kappa_{\lex.R}$ can be understood as follows.

Let us consider the case of $s=0$ and estimate the temperature where the $\kappa_{\lex,R}$ peaks, i.e., where the $\langle w\rangle_R$ peaks. From the numerical result it is clear that as $s\rightarrow 0$ the value of $\langle w\rangle_R$ becomes very close to unity at the peak. By equation (\ref{wR_s0}), when $\langle w\rangle_R$ is close to one we must have $A_{N,k}\equiv\sum_{i=J_k}^N(\hat{\kappa}_i/\kappa_\es)y_i\gg 1$ and $x_N=h\nu_N/kT\ll 1$. In this limit, we can expand the equation and get
\begin{eqnarray}
  \langle w\rangle_R\approx \frac{15}{4\pi^4}\left[\int_{x_N}^\infty f_{dB}(x)dx-\frac{1}{2}\sum_{k=0}^{N-1}\int_{x_{k+1}}^{x_k}A_{N,k}(x)^{-1}f_{dB}(x)dx\right] \;. \label{wR_apr1}
\end{eqnarray}
Since $A_{N,k}(x)\propto (kT)^{-2}(1-e^{-x})x^{-2}$, we can write $A_{N,k}(x)={\cal A}_{N,k} (h\nu_1/kT)^2(1-e^{-x})x^{-2}$, where ${\cal A}_{N,k}$ is independent $x$ and $T$ and defined by
\begin{eqnarray}
  {\cal A}_{N,k}\equiv\frac{1}{\kappa_\es}\frac{\pi e^2}{m_e c}\sum_{j=J_k}^N\frac{n_{1,j}}{\rho_j}\frac{f_{12,j}}{\nu_j}\left(\frac{\nu_j}{\nu_1}\right)^2 \;. \label{calA_Nk}
\end{eqnarray}
Then we have $\langle w\rangle_R\approx 1-\epsilon_1-\epsilon_2$, where $\epsilon_1=(5/4\pi^4)x_N^3$, and
\begin{eqnarray}
  \epsilon_2=\frac{15}{8\pi^4}\left(\frac{h\nu_1}{kT}\right)^{-2}\sum_{k=0}^{N-1}{\cal A}_{N,k}^{-1}\int_{x_{k+1}}^{x_k}\frac{x^6e^{2x}}{(e^x-1)^3}dx \;. \label{ep2}
\end{eqnarray}

Since
\begin{eqnarray}
  \frac{21}{8(\pi^6+945\zeta(5))}\int_0^\infty\frac{x^6e^{2x}}{(e^x-1)^3}dx=1 \;,
\end{eqnarray}
the integral in equation (\ref{ep2}) can be regarded as an average of ${\cal A}_{N,k}^{-1}$. Then we can write
\begin{eqnarray}
  \epsilon_2\approx\frac{5(\pi^6+945\zeta(5))}{7\pi^4}\left(\frac{h\nu_1}{kT}\right)^{-2}\left\langle{\cal A}_{N,k}^{-1}\right\rangle \;, \label{eps2_sol}
\end{eqnarray}
where
\begin{eqnarray}
  \left\langle{\cal A}_{N,k}^{-1}\right\rangle\equiv \frac{21}{8(\pi^6+945\zeta(5))}\sum_{k=0}^{N-1}{\cal A}_{N,k}^{-1}\int_{x_{k+1}}^{x_k}\frac{x^6e^{2x}}{(e^x-1)^3}dx \;.
\end{eqnarray}

We can write $\left\langle{\cal A}_{N,k}^{-1}\right\rangle=f{\cal A}_{N,0}^{-1}$, where ${\cal A}_{N,0}={\cal A}_{N,k=0}$, and $f$ is a number of order unity. By equation (\ref{alp_j_x1}) and the definition of ${\cal A}_{N,k}$ in equation (\ref{calA_Nk}) we have that ${\cal A}_{N,0}=s^{-1}\alpha_{N,1}$. Then, by equation (\ref{eps2_sol}) we have
\begin{eqnarray}
  \epsilon_2\approx f\left(\frac{\chi h\nu_1}{kT}\right)^{-2} \;, 
\end{eqnarray}
where
\begin{eqnarray}
  \chi\equiv\left[\frac{7\pi^4}{5(\pi^6+945\zeta(5))}{\cal A}_{N,0}\right]^{1/2}\approx 0.2650{\cal A}_{N,0}^{1/2} \;. \label{chi}
\end{eqnarray}

Hence, when $s=0$, near the peak of $\langle w\rangle_R$ we have
\begin{eqnarray}
  \langle w\rangle_R\approx 1-\frac{5}{4\pi^4}\left(\frac{h\nu_N}{kT}\right)^3-f\left(\frac{\chi h\nu_1}{kT}\right)^{-2}\;. \label{wR_a1}
\end{eqnarray}

For Nd~II, we have $\chi\approx 166$. When $s=0$ we have $T\approx 9.1\times 10^4\,\K$ at the peak of $\langle w\rangle_R$. Then we get $1-\langle w\rangle_R\approx 7.3\times 10^{-5}+1.3\times 10^{-4}f$. For Fe~IV, we have $\chi\approx 72.7$. When $s=0$ we have $T\approx 1.0\times 10^5\,\K$ at the peak of $\langle w\rangle_R$. Then we get $1-\langle w\rangle_R\approx 6.7\times 10^{-8}+1.2\times 10^{-5}f$. For Fe~III, we have $\chi\approx 129$. When $s=0$ we have $T\approx 1.7\times 10^5\,\K$ at the peak of $\langle w\rangle_R$. Then we get $1-\langle w\rangle_R\approx 9.1\times 10^{-9}+1.7\times 10^{-7}f$. They agree with the numerical results if the value of $f$ is in the range of $\sim 1-2$.

By equation (\ref{wR_a1}) we have
\begin{eqnarray}
  \frac{\partial}{\partial T}\langle w\rangle_R\approx \frac{1}{T}\left[\frac{15}{4\pi^4}\left(\frac{h\nu_N}{kT}\right)^3-2f\left(\frac{\chi h\nu_1}{kT}\right)^{-2}\right]\;, 
\end{eqnarray}
hence $\partial\langle w\rangle_R/\partial T=0$ leads to the temperature at the peak of $\langle w\rangle_R$
\begin{eqnarray}
  T_p\approx\frac{h\nu_1}{k}\left(\frac{15\chi^2}{8\pi^4f}\right)^{1/5}\left(\frac{\nu_N}{\nu_1}\right)^{3/5} \;, \label{Tp_sol}
\end{eqnarray}
in the limit $s\rightarrow 0$.

For Nd~II, we get $T_p\approx 1.8f^{-1/5}h\nu_1/k\approx 7.0f^{-1/5}x_m^{-1}h\nu_1/k$. For Fe~IV, we get $T_p\approx 0.096f^{-1/5}h\nu_1/k\approx 0.4f^{-1/5}x_m^{-1}h\nu_1/k$. For Fe~III, we get $T_p\approx 0.032f^{-1/5}h\nu_1/k\approx 0.1f^{-1/5}x_m^{-1}h\nu_1/k$. These results are fairly consistent with the numerical results in Figs.~\ref{mopc1} and \ref{mopc2}.

Submitting equation (\ref{Tp_sol}) into equation (\ref{wR_a1}), we get
\begin{eqnarray}
  \epsilon_p\equiv 1-\langle w\rangle_R(T=T_p)\approx \frac{5}{3}\left(\frac{15}{8\pi^4}\right)^{2/5}f^{3/5}\left(\frac{\nu_N}{\chi\nu_1}\right)^{6/5}\approx 0.3432f^{3/5}\left(\frac{\nu_N}{\chi\nu_1}\right)^{6/5} \;.
\end{eqnarray}
If $\nu_N/\nu_1\ll 1$, $\chi>1$, and $f\sim 1$, we will then have $\epsilon_p\ll 1$ and $\langle w\rangle_R\approx 1$. As we have seen for the cases of Nd~II, Fe~IV, and Fe~III, these conditions are easily satisfied. For Nd~II, we have $\nu_N/\nu_1\approx 0.339$, $\chi\approx 166$, and then $\epsilon_p\approx 2.0\times 10^{-4} f^{3/5}$. For Fe~IV, we have $\nu_N/\nu_1\approx 0.00429$, $\chi\approx 72.7$, and then $\epsilon_p\approx 2.9\times 10^{-6} f^{3/5}$. For Fe~III, we have $\nu_N/\nu_1\approx 0.000473$, $\chi\approx 129$, and then $\epsilon_p\approx 1.0\times 10^{-7} f^{3/5}$.

\section{Upper Bound and Approximation to the Rosseland Mean Opacity}
\label{apr}

In this section, we derive some useful upper bounds on the Rosseland mean opacity, and some approximation to the Rosseland mean opacity that is easy to use in numerical calculations dealing with the line expansion opacity arising from a huge amount of atomic spectral lines. We note that some interesting bounds on the Rosseland and Planck mean opacity in a general case have been derived in \citet{arm72} and \citet{ber03}.

\subsection{Upper bounds}
\label{upper}

Since by definition we have $w<1$, from equation (\ref{w_cal}) we get
\begin{eqnarray}
  \langle w\rangle_R <\langle w\rangle_{u,1}\equiv\frac{15}{4\pi^4}\int_{x_N}^\infty f_{dB}(x)dx \;. \label{w_lim1}
\end{eqnarray}
As $T\rightarrow 0$, we have $x_N=h\nu_N/kT\rightarrow\infty$ and then $\langle w\rangle_{u,1}\rightarrow (15/4\pi^4)\int_{x_N}^\infty x^4e^{-x}dx=(15/4\pi^4)(x_N^4+4x_N^3+12x_N^2+24x_N+24)e^{-x_N}$. This limit agrees with the limiting value of $\langle w\rangle_R$ as $T\rightarrow 0$ in trend but not in normalization. As $T\rightarrow\infty$, we have $x_N\rightarrow 0$ and then $\langle w\rangle_{u,1}\rightarrow 1$.

A more stringent upper bound on $\langle w\rangle_R$ can be obtained as follows. The integral of $\langle w\rangle_R$ is from $x=x_N$ to $x=x_0=\infty$. In each segment $[x_{k+1},x_k]$ we have $e^{sy_j/2}\le e^{sy_{k+1}/2}$ for $j\ge k+1$. Hence, by equation (\ref{w_cal}) we have
\begin{eqnarray}
  \langle w\rangle_R < \frac{15}{4\pi^4}\sum_{k=0}^{N-1}\int_{x_{k+1}}^{x_k}{\cal C}_{J_k}^{-1}\left[\sum_{j=J_k}^N\left(1-e^{-\tau_j}\right)\exp\left(-\sum_{i=J_k}^{j-1}\tau_i\right)\right]\exp\left[-\frac{s}{2}(1-y_{k+1})\right]f_{dB}dx \;,
\end{eqnarray}
i.e.,
\begin{eqnarray}
  \langle w\rangle_R < \langle w\rangle_{u,2}\equiv\frac{15}{4\pi^4}\sum_{k=0}^{N-1}\int_{x_{k+1}}^{x_k}\frac{1-\exp\left(-\sum_{i=J_k}^N\tau_j\right)}{1-\exp\left(-\sum_{i=J_k}^N\tau_i-s/2\right)}\exp\left[-\frac{s}{2}(1-y_{k+1})\right]f_{dB}dx \label{w_lim2}
\end{eqnarray}
where equation (\ref{sum_id}) has been used. The upper bound $\langle w\rangle_{u,2}$ has a more complex expression than $\langle w\rangle_{u,1}$ but can produce the correct limit of $\langle w\rangle_R$ as $T\rightarrow 0$ and $T\rightarrow\infty$, which can easily be proved.

The $\langle w\rangle_{u,1}$ is independent of $s$. Hence, $\langle w\rangle_{u,1}$ is a universal upper bound for $\langle w\rangle_R$ with any value of $s$. When $T$ is large $\langle w\rangle_{u,1}$ can be far above the value of $\langle w\rangle_R$, since $\langle w\rangle_{u,1}\rightarrow 1$ but $\langle w\rangle_R \propto (kT)^{-2}$ when $T\rightarrow\infty$. From the numerical results in the last section, the limiting solution $\langle w\rangle_R(s=0)$ can be regarded as another universal upper bound on $\langle w\rangle_R$. In contrast, the $\langle w\rangle_{u,2}$ depends on $s$ and can lead to the correct asymptotic values as $T\rightarrow 0$ and $T\rightarrow\infty$. For any nonzero value of $s$, the $\langle w\rangle_{u,2}$ is closer to $\langle w\rangle_R$ than the $\langle w\rangle_{u,1}$.

From $\langle w\rangle_{u,1}$ and $\langle w\rangle_{u,2}$, we can derive another upper bound on $\langle w\rangle_R$:
\begin{eqnarray}
  \langle w\rangle_{u,3}\equiv \langle w\rangle_{N,1}+\langle w\rangle_{1,\infty} \;, \label{w_lim3}
\end{eqnarray}
where
\begin{eqnarray}
  \langle w\rangle_{N,1}\equiv\frac{15}{4\pi^4}\int_{x_N}^{x_1} f_{dB}(x)dx \;, 
\end{eqnarray}
and
\begin{eqnarray}
  \langle w\rangle_{1,\infty}\equiv\frac{15}{4\pi^4}\int_{x_1}^\infty\frac{1-\exp\left(-\sum_{i=1}^N\tau_j\right)}{1-\exp\left(-\sum_{i=1}^N\tau_i-s/2\right)}\exp\left[-\frac{s}{2}(1-y_1)\right]f_{dB}dx \;. \label{w_1inf}
\end{eqnarray}
That is, for the integral from $x_N$ to $x_1$, we take $w=1$ as in the definition of $\langle w\rangle_{u,1}$. For the integral from $x_1$ to $x_0=\infty$, we take the approach used in the definition of $\langle w\rangle_{u,2}$. The $\langle w\rangle_{u,3}$ defined above has an expression simpler than that of $\langle w\rangle_{u,2}$, and can lead to the correct limit expression of $\langle w\rangle_R$ as $T\rightarrow\infty$.

\subsection{Approximation to the Rosseland mean: the case of large $s$}
\label{large_ss}

By equation (\ref{w_cal}), the calculation of $\langle w\rangle_R$ involves a one-dimensional integral across the whole interval of frequency covering all the spectral lines, and a two-dimensional summation over spectral lines. As a result, when the number of spectral lines increases, the computer time spent by the calculation increases rapidly. The numerical calculations presented in Section~\ref{num_result} indicates that, under otherwise the same conditions, the computer time $t_c$ taken by the calculation scales with the total number of spectral lines $N$ involved in the calculation by the relation $t_c\propto N^2$. For instance, Fe~IV has 7.9 times more spectral lines than Nd~II, resulting that the computation of the data for plotting the curve with $s=1$ in the upper panel of Fig.~\ref{mopc2} takes about 62 times more computer time than the computation of the data for plotting the solid curve with $s=1$ in Fig.~\ref{mopc1}. Fe~III has 23 times more spectral lines than Nd~II, resulting that computation of the data for plotting the curve with $s=1$ in the lower panel of Fig.~\ref{mopc2} takes about 450 times more computer time than computation of the data for plotting the solid curve with $s=1$ in Fig.~\ref{mopc1}.

The above fact indicates that when there are a huge amount of spectral lines, direct application of equation (\ref{w_cal}) in the calculation of the Rosseland mean of the line expansion opacity will have a very low efficiency in terms of computer time. Therefore, it is necessary to find some approximation to equation (\ref{w_cal}) that is efficient in computation and in the mean time can lead to a high enough accuracy in the result. In this subsection we will try to derive some approximation to equation (\ref{w_cal}) in the case of $s\gg 1$.

Let us write
\begin{eqnarray}
  \langle w\rangle_R =\frac{15}{4\pi^4}\sum_{k=0}^{N-1}\int_{x_{k+1}}^{x_k}{\cal C}_{k+1}^{-1}(x)F(x)f_{dB}(x)dx \;, \label{wR_F}
\end{eqnarray}
where
\begin{eqnarray}
  F(x) \equiv \sum_{j=k+1}^N\left(1-e^{-\tau_j(x)}\right)\exp\left[-\sum_{i=k+1}^{j-1}\tau_i(x)-\frac{s}{2}(1-y_j(x))\right] \;. \label{F_def}
\end{eqnarray}
When $s$ is large ($s\gg 1$), the dominant contribution to $F(x)$ comes from those lines with $y_j\approx 1$. By definition, we have $y_j=\nu_j^2/\nu^2=x_j^2/x^2$, where $x_j\equiv h\nu_j/kT$. For a given $x$ in the interval $(x_{k+1},x_k)$, we have $y_{k+1}<y_j<1<y_k$ for $j>k+1$. Hence, as $j$ increases from $k+1$, $y_j$ decreases and $1-y_j$ increases. We assume that there exists a $j_m\le N$ such that $(s/2)(1-y_j)$ is sufficiently large when $j>j_m$. In addition, since $\tau_i\propto s$, when $s$ is large $\tau_i$ is also large, the contribution of terms with $j>k+1$ is suppressed quickly as $j$ increases because of the factor $\exp\left(-\sum_{i=k+1}^{j-1}\tau_i\right)$. Hence, when $s$ is large we can truncate the series of $F(x)$ and replace $F(x)$ by
\begin{eqnarray}
  F(x) \approx \sum_{j=k+1}^{j_m}\left(1-e^{-\tau_j(x)}\right)\exp\left[-\sum_{i=k+1}^{j-1}\tau_i(x)-\frac{s}{2}(1-y_j(x))\right] \label{F_apr_ls}
\end{eqnarray}
with negligible errors.

To determine the value of $j_m$, we notice that in the sum of $F(x)$ the ratio of the $j$-th term to the dominant $(k+1)$-th term is given by
\begin{eqnarray}
  \frac{F_j}{F_{k+1}}=\frac{(1-e^{-\tau_j})\exp\left[-\sum_{i=k+1}^{j-1}\tau_i-(s/2)(1-y_j)\right]}{(1-e^{-\tau_{k+1}})\exp\left[-(s/2)(1-y_{k+1})\right]}\approx \exp\left[-\sum_{i=k+1}^{j-1}\tau_i-\frac{s}{2}(y_{k+1}-y_j)\right] \;,
\end{eqnarray}
where we have taken $(1-e^{-\tau_j})/(1-e^{-\tau_{k+1}})\sim 1$. To make the approximation applicable, we want the ratio to be $\ll 1$ when $j$ is larger than some critical value $j_m$. That is, we require that
\begin{eqnarray}
  \gamma(k,j,x)\equiv \sum_{i=k+1}^{j-1}\tau_i+\frac{s}{2}(y_{k+1}-y_j)>\gamma_m \label{jm_eq1}
\end{eqnarray}
when $j=j_m$, where $\gamma_m$ is some positive large number. Since $\gamma(k,j,x)$ is a monotonically increasing function of $j$, for a given pair of $(k,x)$ we have a unique solution of $j_m$, and $\gamma(k,j,x)>\gamma_m$ for all $j\ge j_m$. However, if $\gamma$ is still $<\gamma_m$ when $j=N$, we set $j_m=N$.

The $j_m$ determined above depends on the value of $x$, which causes some inconvenience in the application of the approximation procedure. To eliminate this dependence, we define
\begin{eqnarray}
  \gamma_{k,j}\equiv \sum_{i=k+1}^{j-1}\hat{\tau}_i\left(1-e^{-x_{k+1}}\right)x_k^{-2}+\frac{s}{2}(x^2_{k+1}-x^2_j)x_k^{-2} \;,
\end{eqnarray}
where we have written $\tau_i=\hat{\tau}_i(1-e^{-x})x^{-2}$ and used the relations $y_j=x_j^2/x^2$ and $y_{k+1}=x_{k+1}^2/x^2$. Since $x_{k+1}<x<x_k$, we have $\gamma_{k,j}<\gamma(k,j,x)$. Then, we can defined a $j_m$ by
\begin{eqnarray}
  \gamma_{k,j}>\gamma_m \;. \label{jm_eq2}
\end{eqnarray}
When this equation is satisfied, equation (\ref{jm_eq1}) must also be satisfied. Then we have a unique solution of $j_m$ in each interval $(x_{k+1},x_k)$.

The above procedure does not apply to the interval $(x_1, x_0)$ since then $x_k=x_0=\infty$. For this interval we can just take the sum in $F(x)$ over all spectral lines, i.e., take $j_m=N$.

Since $\hat{\tau}_i\propto s$, when $s$ is large the $\gamma_{k,j}$ defined above increases quickly with $j$ for $j>k+1$. As a result, in reality the solution of $j_m$ can be very close to $k+1$, i.e., $j_m=k+\mbox{a few}$. This fact makes the approximation given by equations (\ref{wR_F}) and (\ref{F_apr_ls}) to be very efficient in evaluation when $s$ is large, and the results are very accurate so long as $s$ is not too small.

According to the above description, the accuracy in the $\langle w\rangle_R$ evaluated with this approach is determined by the value of $\gamma_m$. If $\gamma_m$ is reasonably large, we would get a $\langle w\rangle_R$ with a sufficiently high accuracy, no matter $s$ is large or small. The effect of the approximation is in saving the computer time when $s$ is large, not in affecting the accuracy of $\langle w\rangle_R$. This is because that by the definition of the approach, if terms in the sum of $F(x)$ do not drop rapidly as $j$ increases (caused by, for instance, a small value of $s$), we will have $j_m=N$ then the result is exact without saving the computer time.

\begin{figure}[ht!]
\centering{\includegraphics[angle=0,scale=0.7]{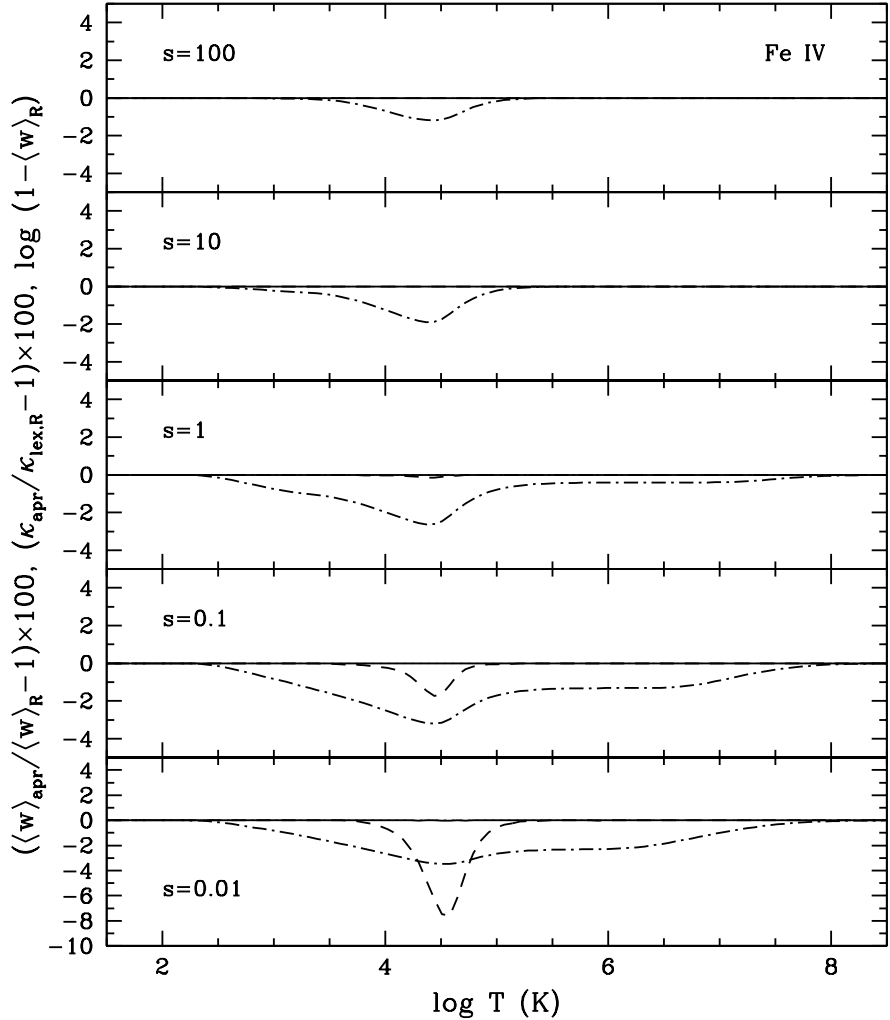}}
\caption{Fractional errors in the $\langle w\rangle_R$ (solid curve) and $\kappa_{\lex,R}$ (dashed curve) calculated with the approximate solution defined by equations (\ref{wR_F}) and (\ref{F_apr_ls}) relative to the exact solution, for the case of Fe IV. The exact solutions of $1-\langle w\rangle_R$ are shown with dash-dotted curves. As explained in the text, when $s$ is not too large $\langle w\rangle_R$ can be very close to 1 near the peak so that $1-\langle w\rangle_R$ is very small. As a result, the fractional errors in $\kappa_{\lex,R}$ evaluated by equation (\ref{kap_wr}) are amplified relative to the fractional errors in $\langle w\rangle_R$. However, for $s\ge 1$, we always have $\eta_\kappa\equiv\left|\kappa_{\rm apr}/\kappa_{\rm lex,R}-1\right|<0.2\%$. Even for the case of $s=0.1$, we still have $\eta<2\%$. When $s>10$, we have $\eta<0.01\%$.
}
\label{mopc2_apr2}
\end{figure}

\begin{figure}[ht!]
\centering{\includegraphics[angle=0,scale=0.7]{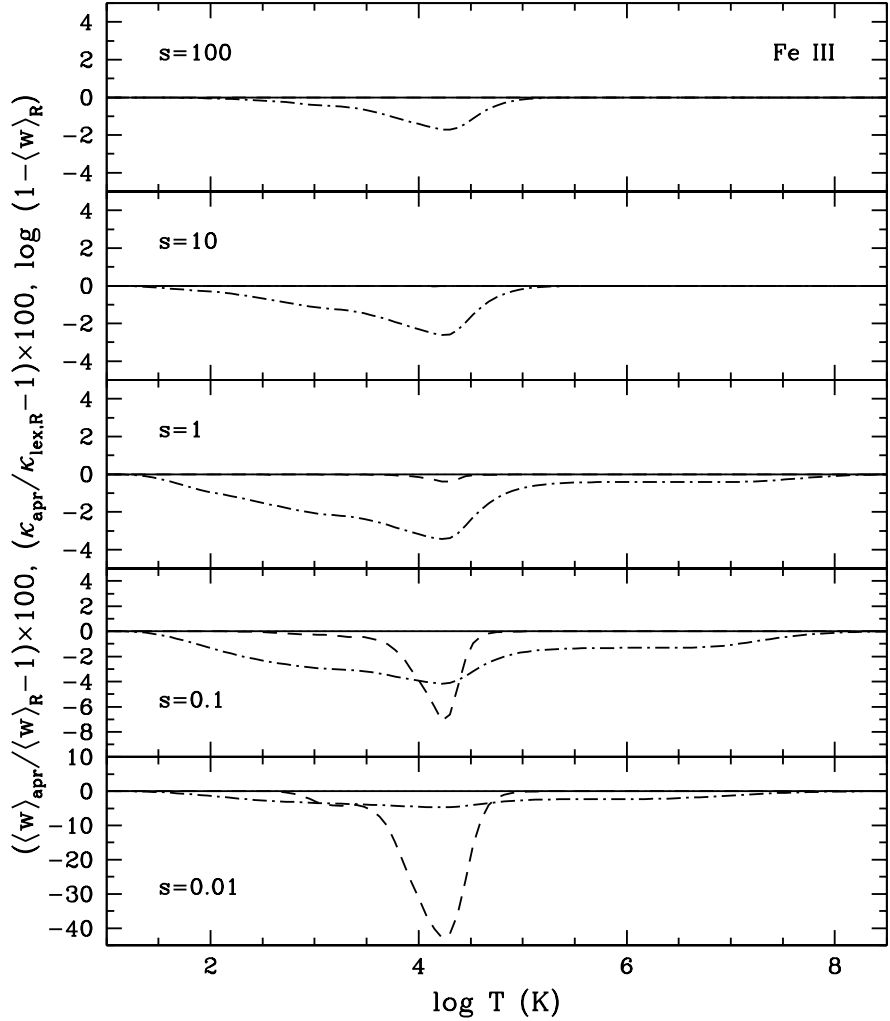}}
\caption{Similar to Fig.~\ref{mopc2_apr2} but for Fe~III. For $s\ge 1$, we always have $\eta_\kappa=\left|\kappa_{\rm apr}/\kappa_{\rm lex,R}-1\right|<0.4\%$. Even for the case of $s=0.1$, we still have $\eta<8\%$. When $s>10$, we have $\eta<0.01\%$.
}
\label{mopc3_apr2}
\end{figure}

Hence, if we choose a reasonably large value of $\gamma_m$, the accuracy in $\langle w\rangle_R$ is guaranteed by the approximation. For instance, if we choose $\gamma_m=10$, we can get $\langle w\rangle_R$ with a fractional error $\la e^{-10}\approx 4.5\times 10^{-5}$. This is confirmed by numerical calculations for the cases of Nd~II, Fe~IV, and Fe~III. However, this does not mean that we can also obtain the $\kappa_{\lex,R}$ with the same accuracy. By equation (\ref{kap_wr}), the error in $\kappa_{\lex,R}$ is related to the error in $\langle w\rangle_R$ by
\begin{eqnarray}
  \frac{\delta\kappa_{\lex,R}}{\kappa_{\lex,R}}\approx\frac{\delta\langle w\rangle_R}{\langle w\rangle_R}\frac{1}{1-\langle w\rangle_R} \;. \label{dkap_kap}
\end{eqnarray}
If $\langle w\rangle_R$ is very close to unity, we will have $\delta\kappa_{\lex,R}/\kappa_{\lex,R}\gg \delta\langle w\rangle_R/\langle w\rangle_R$.

From Figs.~\ref{mopc1} and \ref{mopc2} we see that, for the cases that we have calculated, $\langle w\rangle_R$ can indeed be very close to one (then $\kappa_{\lex,R}\gg\kappa_\es$) when $s$ is $\la 100$, at least near the peak of $\langle w\rangle_R$. This fact indicates that when $s$ is not very large the $\kappa_{\lex,R}$ evaluated with the approximate approach can indeed suffer an appreciable error even though the $\langle w\rangle_R$ is evaluated very accurately.

We have applied the approximation approach defined above to calculate the $\langle w\rangle_R$ and $\kappa_{\lex,R}$ for Nd~II, Fe~IV, and Fe~III for $s\ge 0.01$, and compared the result to that obtained with the exact formulae in Section~\ref{num_result}. We find that, as expected, in all cases the fractional error in $\langle w\rangle_R$ is $<10^{-4.5}$. For the case of Nd~II, the fractional error in $\kappa_{\lex,R}$ is always $<1\%$, since $1-\langle w\rangle_R\ga 10^{-2.3}$ always. For the case of Fe~IV, the error in $\kappa_{\lex,R}$ becomes noticible ($>1\%$) near the peak only for $s<1$. However, even for $s=0.01$ the error in $\kappa_{\lex,R}$ is still acceptible ($<10\%$), as shown in Fig.~\ref{mopc2_apr2}. The case of Fe~III is similar to that of Fe~IV for $s\ga 1$. Although for $s=0.1$ we can still get an error $<10\%$ in $\kappa_{\lex,R}$, for $s=0.01$ the error in $\kappa_{\lex,R}$ is about $43\%$ at the peak (Fig.~\ref{mopc3_apr2}).

When $s=0.01$, we have $1-\langle w\rangle_R\approx 3.3\times 10^{-4}$ for Fe~IV, and $1-\langle w\rangle_R\approx 1.9\times 10^{-5}$ for Fe~III at the peak of $\langle w\rangle_R$. The $1-\langle w\rangle_R$ for Fe~III is 17 times smaller than that for Fe~IV, explaining why the error in $\kappa_{\lex,R}$ for Fe~III is much larger than that for Fe~IV ($43\%$ vs. $8\%$ at peak). For comparison, when $s=0.01$ we have $1-\langle w\rangle_R\approx 0.0049$ for Nd~II, about 15 times that for Fe~IV, explaining why the error in $\kappa_{\lex,R}$ is $<1\%$ for Nd~II.

Overall, the approximation procedure defined above seems to work very well for cases with $s\ga 1$, considerably saving the time in computation and preserving a good accuracy in both $\langle w\rangle_R$ and $\kappa_{\lex,R}$. The more spectral lines, the more time the computation can save, since then more lines are useless in computation of the integral in a given frequency interval. We can take the case of $s=1$ and $\gamma_m=10$ as an example. For Nd~II, computation with the approximation approach takes about one-fourth the time taken by the exact formula. For Fe~IV, which has about 8 times more spectral lines than Nd~II, computation with the approximation approach takes about one-twelveth the time taken by the exact formula. For Fe~III, which has about 23 times more spectral lines than Nd~II, computation with the approximation approach takes about one-twentieth the time taken by the exact formula. This fact means that the more spectral lines the better the approximation works.

Clearly, the above approximation approach can also be applied to the calculation of the monochromatic line expansion opacity. That is, to calculate the $w(\nu)$ and $\kappa_\lex(\nu)$, the $\sum_{j=J}^N$ in equation (\ref{kap_exp1b}) can be replaced by $\sum_{j=J}^{j_m}$, but here the $j_m$ is determined by equation (\ref{jm_eq1}).

\subsection{Asymptotic solution of $\langle w\rangle_R$ in the limit of $s\rightarrow\infty$}
\label{s_inf}

By equation~(\ref{F_def}), when $s$ is large (then $\tau_j$'s are also large) only the term with $j=k+1$ makes nonnegligible contribution to $F(x)$. Then we have
\begin{eqnarray}
  F(x)\approx\left(1-e^{-\tau_{k+1}}\right)\exp\left[-\frac{s}{2}\left(1-y_{k+1}\right)\right]\approx\exp\left[-\frac{s}{2}\left(1-\frac{x_{k+1}^2}{x^2}\right)\right] \;. \label{F_apr2}
\end{eqnarray}
When $s\rightarrow\infty$, $F(x)$ behaves like a Dirac $\delta$-function since for any $x>x_{k+1}$ we should have $\lim_{s\rightarrow\infty}F(x)=0$. In fact, if we define
\begin{eqnarray}
  \Phi(x,x_{k+1})=\frac{s}{x_{k+1}}\exp\left[-\frac{s}{2}\left(1-\frac{x_{k+1}^2}{x^2}\right)\right]\vartheta(x-x_{k+1})
\end{eqnarray}
where $\vartheta(x)$ is the Heaviside step function, it can be proved that
\begin{eqnarray}
  \lim_{s\rightarrow\infty}\Phi(x,x_{k+1})=\delta(x-x_{k+1}) \;. \label{phi_del}
\end{eqnarray}

The proof of equation (\ref{phi_del}) is simple. First, it is easy to see that when $x\neq x_{k+1}$ we have $\lim_{s\rightarrow\infty}\Phi(x,x_{k+1})=0$. Next, we need to prove
\begin{eqnarray}
  \lim_{s\rightarrow\infty}\int_0^{x_1}\Phi(x,x_{k+1})dx=1 \;, \hspace{1cm} \mbox{for any $x_1>x_{k+1}$} \;. \label{int_Phi}
\end{eqnarray}
It is easy to derive that
\begin{eqnarray}
  \int_0^{x_1}\Phi(x,x_{k+1})dx = 2s\sqrt{\frac{s}{2}}\left\{D\left(\sqrt{\frac{s}{2}}\right)-\frac{1}{2}\sqrt{\frac{2}{s}}-e^{-\frac{s}{2}\left(1-\frac{x_{k+1}^2}{x_1^2}\right)}\left[D\left(\sqrt{\frac{s}{2}}\frac{x_{k+1}}{x_1}\right)-\frac{1}{2}\sqrt{\frac{2}{s}}\frac{x_1}{x_{k+1}}\right]\right\} \;, \label{Phi_int1}
\end{eqnarray}
where $D(x)$ is the Dawson function. When $x$ is large we have $D(x)=1/(2x)+1/(4x^3)+{\cal O}(x^{-5})$. Hence, when $s$ is large we have
\begin{eqnarray}
  \int_0^{x_1}\Phi(x,x_{k+1})dx =1-\frac{x_1^3}{x_{k+1}^3}e^{-\frac{s}{2}\left(1-\frac{x_{k+1}^2}{x_1^2}\right)}+{\cal O}\left(s^{-5/2}\right) \;, \label{Phi_int2}
\end{eqnarray}
then equation (\ref{int_Phi}) is proved.

Hence, we have
\begin{eqnarray}
  \lim_{s\rightarrow\infty}sF(x)=x_{k+1}\delta(x-x_{k+1}) \;. \label{sF_lim}
\end{eqnarray}

When $s$ is large we have ${\cal C}_{k+1}\approx 1$. Then, by equations (\ref{wR_F}) and (\ref{sF_lim}), when $s$ is large we have
\begin{eqnarray}
  \langle w\rangle_R \approx \frac{15}{4\pi^4s}\sum_{k=0}^{N-1}x_{k+1}\int_{x_{k+1}-\epsilon}^{x_k}f_{dB}(x)\delta(x-x_{k+1})dx=\frac{A}{s} \;, \label{wR_ls}
\end{eqnarray}
where
\begin{eqnarray}
  A \equiv\frac{15}{4\pi^4}\sum_{k=0}^{N-1}x_{k+1}f_{dB}(x_{k+1}) \;. \label{A_def}
\end{eqnarray}
Thus, we have $\langle w\rangle_R\propto s^{-1}$ when $s$ is large.

Although going from equation (\ref{Phi_int1}) to equation (\ref{Phi_int2}) and then to equation (\ref{sF_lim}) requires only $s\gg 1$, the approximation of $F(x)$ in equation (\ref{F_apr2}) requires that all $\tau_i\gg 1$. The condition $\tau_i\gg 1$ is usually more stringent than the condition $s\gg 1$, since some lines can be very weak. For instance, from Fig.~\ref{mopc3_s} we see that $\langle w\rangle_R$ (and hence $\kappa_{\lex,R}$) approaches the asymptotic form $\langle w\rangle_R\propto s^{-1}$ only when $s$ is $\ga 10^6$ for the case of Fe~III.

When $T\rightarrow\infty$, we have $x_{k+1}=h\nu_{k+1}/kT\rightarrow 0$ and $f_{dB}(x_{k+1})\approx x_{k+1}^2$. Then we have $A \approx(15/4\pi^4)\sum_{k=0}^{N-1}x_{k+1}^3$, and
\begin{eqnarray}
  \langle w\rangle_R \approx\frac{15}{4\pi^4 s}\sum_{k=0}^{N-1}x_{k+1}^3\propto s^{-1}(kT)^{-3} \;. \label{wR_sT}
\end{eqnarray}
Note, the asymptotic solution given by equation (\ref{wR_sT}) does not agree with the solution (\ref{wR_Ti1}) that we have derived before. As $s\rightarrow\infty$ we have $e^{-s/2}\rightarrow 0$, hence the asymptotic solution in equation (\ref{wR_sT}) corresponds to the region after the peak but before the flat part on the $\langle w\rangle_R$-$T$ curve. In contrast, the asymptotic solution (\ref{wR_Ti1}) applies when $kT\gg\alpha_{N,1}^{1/2}h\nu_1$. As $s\rightarrow\infty$ we have $\alpha_{N,1}\rightarrow\infty$, then the asymptotic solution (\ref{wR_Ti1}) does not apply.

From $\langle w\rangle_R=e^{-s/2}$ we can solve for the temperature $T=T_s$ where the flat part on the $\langle w\rangle_R$-$T$ curve begins. The solution is
\begin{eqnarray}
  T_s=\frac{h\nu_1}{k}\frac{e^{s/6}}{s^{1/3}}\left[\frac{15}{4\pi^4}\sum_{k=0}^{N-1}\left(\frac{\nu_{k+1}}{\nu_1}\right)^3\right]^{1/3} \;.
\end{eqnarray}
As $s\rightarrow\infty$ we have $T_s\rightarrow\infty$, so the flat part and the solution after the flat part are unimportant when $s$ is very large.

When $T\rightarrow 0$, we have $x_{k+1}\rightarrow\infty$ and $f_{dB}(x_{k+1})\approx x_{k+1}^4e^{-x_{k+1}}$. Then we have $A \approx(15/4\pi^4)\sum_{k=0}^{N-1}x_{k+1}^5e^{-x_{k+1}}$. Since for $k+1<N$ we have $(x_{k+1}/x_N)^5e^{-(x_{k+1}-x_N)}=(\nu_{k+1}/\nu_N)^5e^{-h(\nu_{k+1}-\nu_N)/kT}\rightarrow 0$ as $T\rightarrow 0$, we have $A \approx(15/4\pi^4)x_N^5e^{-x_N}$ as $T\rightarrow 0$. This result agrees with the asymptotic solution (\ref{w_T_lim1a}) in the limit $s\rightarrow\infty$. Therefore, the approximate solution defined by equations (\ref{wR_ls}) and (\ref{A_def}) leads to the correct asymptotic solution as $T\rightarrow 0$ and $T\rightarrow\infty$.

If we define the $s$-parameter at any time $\eta$ by $s(\eta)=\kappa_\es \rho c\eta$, by $\rho=\rho_0(\eta/\eta_0)^{-3}$ we get
\begin{eqnarray}
  s(\eta)=s_0\left(\frac{\rho}{\rho_0}\right)^{2/3} \;, \label{s_rho}
\end{eqnarray}
where $s_0\equiv\kappa_\es\rho_0c\eta_0$. Then, by equation (\ref{wR_ls}) we get
\begin{eqnarray}
  \langle w\rangle_R =\frac{A}{s_0}\left(\frac{\rho}{\rho_0}\right)^{-2/3}\propto\left(\frac{\rho}{\rho_0}\right)^{-2/3} \;, \hspace{1cm} \mbox{as $\rho\rightarrow\infty$}  \;. \label{wR_ls2}
\end{eqnarray}

\subsection{Approximation to the Rosseland mean: the case of small $s$}
\label{small_ss}

Since $0<y_j<1$, when $s\ll 1$ we have $s(1-y_j)\ll 1$ and $\exp[-(s/2)(1-y_j)]\approx 1$. Then, by equation (\ref{w_cal}) we get
\begin{eqnarray}
  \langle w\rangle_R &\approx& \frac{15}{4\pi^4}\sum_{k=0}^{N-1}\int_{x_{k+1}}^{x_k}{\cal C}_{J_k}^{-1}\left[\sum_{j=J_k}^N\left(1-e^{-\tau_j}\right)\exp\left(-\sum_{i=J_k}^{j-1}\tau_i\right)\right]f_{dB}dx \nonumber\\
  &=& \frac{15}{4\pi^4}\sum_{k=0}^{N-1}\int_{x_{k+1}}^{x_k}\frac{1-\exp\left(-\sum_{i=J_k}^N\tau_i\right)}{1-\exp\left(-\sum_{i=J_k}^N\tau_i-s/2\right)}f_{dB}dx \;. \label{w_u1a}
\end{eqnarray}
As $s\rightarrow 0$, the $\langle w\rangle_R$ approaches the expression in equation (\ref{wR_s0}). For small but nonzero $s$ equation (\ref{w_u1a}) is a better expression, which can be different from that in equation (\ref{wR_s0}) when $\sum_{i=J_k}^N\tau_i$ is not small.

As demonstrated in Section~\ref{large_ss}, an accurate evaluation of $\langle w\rangle_R$ does not necessarily lead to an accurate evaluation of $\kappa_{\lex,R}$, since $\langle w\rangle_R$ can be very close to unity. When $s$ is small, the error in $\langle w\rangle_R$ arising from ignorance of the effect of $s$ in the factor $\exp[-(s/2)(1-y_j)]$ can be estimated by $\delta\langle w\rangle_R\sim (s/2)\langle w\rangle_R$. Then, by equation (\ref{dkap_kap}) we have
\begin{eqnarray}
  \frac{\delta\kappa_{\lex,R}}{\kappa_{\lex,R}}\sim\frac{s}{2}\frac{1}{1-\langle w\rangle_R} \;. \label{dkap_kap2}
\end{eqnarray}
When $\langle w\rangle_R$ is close to one, we have $0<1-\langle w\rangle_R\ll 1$ and then $\delta\kappa_{\lex,R}/\kappa_{\lex,R}\gg \delta\langle w\rangle_R/\langle w\rangle_R\sim s/2$. To have $\delta\kappa_{\lex,R}/\kappa_{\lex,R}<\varepsilon_\kappa$ where $\varepsilon_\kappa$ is a small number, we must have $s<2\varepsilon_\kappa(1-\langle w\rangle_R)$. For instance, if $1-\langle w\rangle_R=10^{-4}$, to get $\delta\kappa_{\lex,R}/\kappa_{\lex,R}<\varepsilon_\kappa=5\%$ we need to have $s<10^{-5}$.

\begin{figure}[ht!]
\centering{\includegraphics[angle=0,scale=0.7]{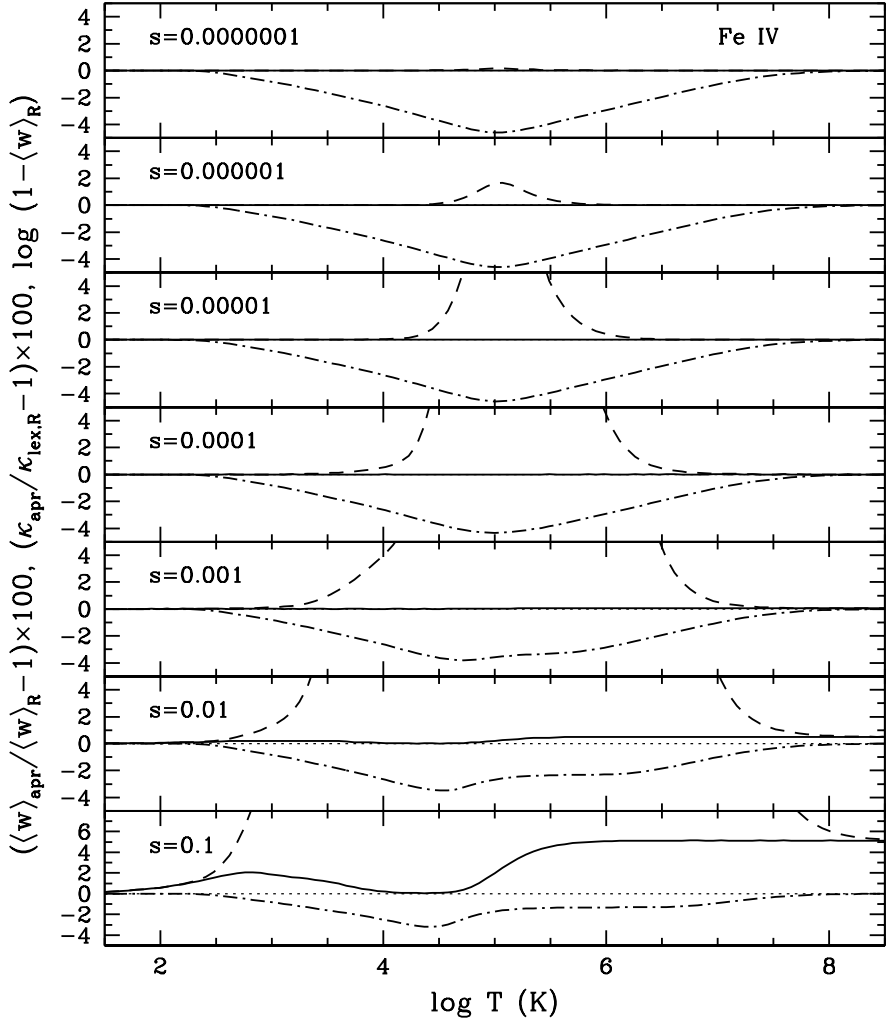}}
\caption{Fractional errors in the $\langle w\rangle_R$ (solid curve) and $\kappa_{\lex,R}$ (dashed curve) evaluated with the approximate solution in equation (\ref{w_u1a}) relative to the exact solution, for the case of Fe IV with small $s$. The exact solutions of $1-\langle w\rangle_R$ are shown with dash-dotted lines. As explained in the text, when $s$ is small $\langle w\rangle_R$ can be very close to 1 so that $1-\langle w\rangle_R$ is very small. As a result, the fractional errors in $\kappa_{\lex,R}$ evaluated by equation (\ref{kap_wr}) are amplified relative to the fractional errors in $\langle w\rangle_R$. The dotted line shows the location of zero of the vertical axis.
}
\label{mopc2_apr1}
\end{figure}

We take Fe~IV as an example to calculate the $\langle w\rangle_R$ and $\langle\kappa\rangle_{\lex,R}$ with the approximate solution in equation (\ref{w_u1a}) for a number of small $s$, then compare the result to that calculated with the exact solution given by equation (\ref{w_cal}). The fractional errors in the approximate solutions relative to the exact solutions are shown in Fig.~\ref{mopc2_apr1}. We see that, for $s\la 0.1$ we can get very accurate $\langle w\rangle_R$ with the approximate solution, but $\langle\kappa\rangle_{\lex,R}$ with an acceptable accuracy can only be obtained for $s<10^{-5}$. This agrees with the above analysis. For instance, when $s=10^{-5}$ we have $1-\langle w\rangle_R\approx 2.5\times 10^{-5}$ at the peak of $\langle w\rangle_R$. Then by equation (\ref{dkap_kap2}) we have $\delta\langle\kappa\rangle_{\lex,R}/\langle\kappa\rangle_{\lex,R}\sim 20\%$. This estimated value agrees with the numerical result shown in the figure. Since $\delta\langle w\rangle_R/\langle w\rangle_R\sim s/2$, we can get $\langle w\rangle_R$ with a fractional error $<5\%$ for $s<0.1$.

The above results indicate that when $s$ is small so that $\langle w\rangle_R$ is close to one, it is very challenging to get an accurate evaluation of the Rosseland mean opacity $\kappa_{\lex,R}$ numerically. The closer to one the $\langle w\rangle_R$ is, the harder it is to get the $\kappa_{\lex,R}$ accurately.

\section{Discussion: The Case of a Nonuniform Sphere}
\label{disuss}

Here we consider a spherically symmetric sphere with a nonuniform mass density as a generalization of the spherical and uniform model that has been considered so far. Like often did in the investigation of supernovae and neutron star mergers, we assume that in a short time after the explosion (a few days for a supernova and a few seconds for a merger) the ejecta settles into a free and homologous expansion with the expansion velocity proportional to the distance to the sphere center \citep{rop05,lip15,rad18}. The expansion is spherical and uniform,\footnote{That is, the gradient of the 4-velocity of the spherical fluid has only the expansion component, no shear, vorticity, and acceleration. In a frame comoving with the fluid motion the expansion does not depend on position and direction.} so that the metric inside the sphere is still given by the Milne metric. Since the gravity of the matter is negligible, the nonuniformity in the mass density has no effect on the spacetime metric. In the homologous approximation, the mass contained in a spherical shell defined by radius $\xi$ and $\xi+d\xi$ is conserved, from which we derive that $\rho\propto\eta^{-3}$. Hence, we have the relation
\begin{eqnarray}
  \rho(\xi,\eta)=\rho_0(\xi)\left(\frac{\eta}{\eta_0}\right)^{-3} \;, \hspace{1cm}
  \rho_0(\xi)\equiv\rho(\xi,\eta_0) \;. \label{rho_eta}
\end{eqnarray}
In the case of a uniform sphere $\rho_0$ does not depend on $\xi$ hence $\rho_0$ is a constant. But here, where the ejecta is assumed to be a spherical but nonuniform, $\rho_0$ is a function of $\xi$.

Note, in the Newtonian limit we have $\eta= t$ and $\xi=\beta=v/c$, and then equation (\ref{rho_eta}) becomes the standard Newtonian one for the homologous expansion as often seen in the literature.

The evolution of the photon frequency is determined by the kinematics of the sphere, or equivalently, by the spacetime metric inside the sphere. Since the spacetime inside the sphere is still described by the Milne metric, like in the case of an expanding sphere with a uniform mass density, we have
\begin{eqnarray}
  \nu=\nu_0\left(\frac{\eta}{\eta_0}\right)^{-1} \;, \label{nu_eta}
\end{eqnarray}
i.e., the frequency of the photon is redshifted by the uniform expansion of the sphere.

Following the procedure in Section~\ref{l_opc}, we can derive that the optical depth of a spectral line of frequency $\nu_l$ to a photon of frequency $\nu_0>\nu_l$ is still given by equations (\ref{tau_l0}) and (\ref{hkap_l}), however where $\rho_l$, $n_{1l}$ and $T_l$ should be evaluated at the place and time where and when the photon encounters the line. Consider a photon moving radially along a null geodesics. Then we have $\xi=\xi_0\pm\ln(\eta/\eta_0)$ along the trajectory of the photon, where $(\xi_0,\eta_0)$ are the initial coordinates of the photon. Thus, at time $\eta$, the mass density at the location of the photon is
\begin{eqnarray}
  \rho(\xi,\eta)=\rho_0\left(\xi=\xi_0\pm\ln\frac{\eta}{\eta_0}\right)\left(\frac{\eta}{\eta_0}\right)^{-3} \;.
\end{eqnarray}
Hence, in equations (\ref{tau_l0}) and (\ref{hkap_l}) we should have $\rho_l=\rho_0\left(\xi_l\right)(\eta_l/\eta_0)^{-3}$ where $\xi_l=\xi_0\pm\ln(\eta_l/\eta_0)$. Similarly, we should have $n_{1l}=n_1(\xi_l,\eta_l)$ and $T_l=T(\xi_l,\eta_l)$.

If we define a position-dependent $s$-parameter by $s(\xi_0)\equiv\kappa_\es\rho_0(\xi_0)c\eta_0$, we have $\rho c\eta=(s/\kappa_\es)(\nu/\nu_0)^2\rho_0(\xi)/\rho_0(\xi_0)$. Then, generalization of equation (\ref{tau_l}) to the case of a sphere with a nonuniform mass density leads to
\begin{eqnarray}
  \tau_l(\nu_0)=s\frac{\hat{\kappa}_l}{\kappa_\es}\frac{\nu_l^2}{\nu_0^2}\frac{\rho_0(\xi_l)}{\rho_0(\xi_0)}\vartheta(\nu_0-\nu_l) \;. \label{tau_lg}
\end{eqnarray}

The mean free path of a photon propagating in such a nonuniform sphere is calculated by equation (\ref{l_bar}), where the integral is defined along the light ray. Since the mass density is a function of radius, the optical depth depends on the direction of the light ray and so does the integral for the mean free path, unless the light ray starts from the center of the sphere. To obtain a mean free path (and hence an averaged opacity) that is not a function of direction, we can average the optical depth over directions and then calculated the mean free path. That is, we can define
\begin{eqnarray}
  \bar{\tau}(l)\equiv\frac{1}{4\pi}\int\tau(l,\theta,\phi)\sin\theta d\theta d\phi \;,
\end{eqnarray}
where the angles $(\theta,\phi)$ define the direction of the light ray. Then we define the mean free path by
\begin{eqnarray}
  \bar{X}\equiv \int_0^\infty e^{-(\bar{\tau}_\nu+\bar{\tau}_s)}\left(\frac{\nu}{\nu_0}\right)^{3}dl \;.
\end{eqnarray}

In the treatment of \citet{kar77} it has been assumed that the variation in the mass density of the sphere is negligible within a distance of the mean free path of the photon and then the mean free path is independent of the direction of the photon motion. This assumption is valid only if the size of the sphere is much larger than the mean free path of the photon, i.e., only if the sphere is optically thick so that the diffusion approximation holds.

As we have have seen in Section~\ref{l_exp_opc}, the mean free path of the universal expansion is $X_H=c\eta_0/2$ (independent of direction), corresponding to $\Delta\eta=\eta_0/2$. Hence, after a photon moving a distance corresponding to the mean free path of the universal expansion, the comoving time is $\eta=\eta_0+\Delta\eta=3\eta_0/2$ and the frequency of the photon is $\nu=2\nu_0/3$. By the geodesic equation along the radial direction, at $\eta=3\eta_0/2$ we have $\xi=\xi_0\pm\ln(3/2)$. Hence, the frequency of a photon varies on a time scale $\Delta\eta_\nu\approx\eta_0/2$, or equivalently, on a spatial scale $\Delta\xi_\nu\approx\ln(3/2)=0.405$. Note, on a hypersurface of $\eta=\mbox{const}$ the proper distance along the radial direction is $\Delta l=c\eta\Delta\xi$. Hence, $\xi$ measures the proper distance in the comoving frame of the sphere. At the surface of the sphere we have $\xi=\xi_R=\arctanh\beta_R$. In the Newtonian limit we have $\xi_R\approx \beta_R\ll 1$. In the ultra-relativistic limit we have $\xi_R\approx \ln(2\gamma_R)$, where $\gamma_R=(1-\beta_R^2)^{-1/2}\gg 1$. Therefore, in the Newtonian limit the variation in the frequency of photon along a light ray is small,\footnote{This is the reason for the validity of the equation (3) in \citet{kar77}.} while in a relativistic case the variation in the frequency of a photon as it travels through the sphere is appreciable.

The gradient of the mass density with respect to the spatial coordinate in the comoving frame is determined by the amplitude of $d\ln\rho_0/d\xi$. Since $d\rho_0/d\xi\sim \rho_0/\xi_R$, we have $d\ln\rho_0/d\xi\sim \xi_R^{-1}$. This relation seems to indicate that the spatial gradient of the mass density of the sphere is less important in a relativistic case than in a Newtonian case.

\section{Summary and Conclusions}
\label{sum}

We have derived the line expansion opacity arising from spectral lines of atomic elements in a relativistically expanding medium. For simplicity, it has been assumed that the medium is a uniformly expanding sphere with a uniform mass density, although generalization to a sphere with a nonuniform mass density has also been discussed. The derived line expansion opacity is given by equations (\ref{kap_exp1a})--(\ref{cal_CJ}), where the enhancement factor $w(\nu_0)$ defines the fraction of the line expansion opacity in the total opacity. The Rosseland mean of the line expansion opacity is given by equations (\ref{ros_w}) and (\ref{w_cal})--(\ref{h_kap_j}).

To derive the line expansion opacity we have introduced a mean free path defined by equation (\ref{l_bar}), which includes the contribution from a universal expansion opacity. The universal expansion opacity defined by equation (\ref{kap_H}) is attributed to the Doppler effect arising from the uniform expansion of the sphere. In the treatment of \citet{kar77}, which was based on the Newtonian theory, the universal expansion opacity is not included in the definition for the mean free path. In the relativistic treatment inclusion of the universal expansion opacity is necessary to guarantee the convergence of the integral for the mean free path. Hence, equation (\ref{l_bar}) can be considered as relativistic generalization of the mean free path equation adopted in \citet{kar77}.

An important parameter in the derived line expansion opacity is the expansion parameter $s$, which was originally introduced by \citet{kar77}. The $s$-parameter is related to the optical depth of the matter to electron scattering (eq.~\ref{tau_es_beta}), or equivalently, to the mass density of the matter (eq.~\ref{s_rho}). As $s\rightarrow\infty$, the line expansion opacity derived in this paper approaches that derived by \citet{kar77} in the Newtonian limit. When $s$ is small or around unity, the result given by our formulae is more accurate than that given by the formulae of \citet{kar77}. For instance, when $s\rightarrow 0$ our formulae lead to a finite and nonzero line expansion opacity, while the formulae of \citet{kar77} lead to a zero line expansion opacity (Fig.~\ref{mopc1}).

Hence, the line expansion opacity derived by \citet{kar77} is valid only when the matter is opaque to electron scattering. When the matter is transparent to electron scattering, or, during a stage when the matter transits from an opaque state to a transparent state with respect to electron scattering, the formulae derived in this paper for the line expansion opacity must be used to get the correct result.

In the case of neutron star mergers, it is expected that the merger ejecta starts from an optically thick state and becomes optically thin to electron scattering in a day to a couple of days after the merger \citep{li98}. This has been confirmed by the extensive observation on the optical transient associated with GW170817/GRB170817A---the first neutron star merger that has ever been discovered in both gravitational and electromagnetic wave bands \citep{eva17,mcc17,pia17,buc18}. Understanding the process of transition from an optically thick phase to an optically thin phase is critical in the study of neutron star mergers, where the line expansion opacity derived in this paper should have important applications.

In fact, we expect that the effect of the line expansion opacity is most important in the transition phase with $s\sim 1$ during the evolution of a transient source. As the results in this paper have shown, when $s\rightarrow\infty$ we have $\kappa_{\lex,R}\propto s^{-1}$, indicating that when $s\gg 1$ the opacity is dominated by that arising from electron scattering. When $s\rightarrow 0$ we have $\kappa_{\lex,R}\rightarrow{\cal O}(s^0)$, indicating that when $s\ll 1$ the optical depth arising from the line expansion opacity approaches zero due to the fast drop in the mass density (see Fig.~\ref{mopc3_s}).

We have also shown that the simplified equation for the line expansion opacity, which is often adopted in the numerical works in the literature, i.e., equation (\ref{simp_kap}), can be derived from our generalized formulae under the assumption that spectral lines are densely and uniformly distributed in a small range of wavelength. When this assumption is violated the simplified equation (\ref{simp_kap}) does not apply in principle.

Besides the atomic compositions and the parameters of spectral lines, the Rosseland mean of the line expansion opacity in an expanding sphere, $\kappa_{\lex,R}$, depends on two critical parameters: the temperature $T$, and the expansion parameter $s$ (or, equivalently, the mass density $\rho$). The asymptotic trend of $\kappa_{\lex,R}$ (which is also the trend of $\langle w\rangle_R$---the Rosseland mean of the enhancement factor) with respect to $T$ and $s$ can be summarized as follows: (1) When $s$ is fixed but $T\rightarrow 0$, we have $\kappa_{\lex,R}\propto (kT)^{-4}e^{-h\nu_N/kT}$; when $s$ is fixed but $T\rightarrow\infty$, we have $\kappa_{\lex,R}\propto (kT)^{-2}$ (Fig.~\ref{wr1}). (2) When $T$ is fixed but $s\rightarrow 0$, we have $\kappa_{\lex,R}\propto s^0$; when $T$ is fixed but $s\rightarrow\infty$, we have $\kappa_{\lex,R}\propto s^{-1}$. As $s\rightarrow\infty$ the flat part on the $\kappa_{\lex,R}$-$T$ curve moves to infinity, then as $T\rightarrow\infty$ we get $\kappa_{\lex,R}\propto (kT)^{-3}$ (Sec.~\ref{s_inf}). Therefore, in the very early stage of the merger or explosion when both the mass density and the temperature are very high, we have $\kappa_{\lex,R}\propto s^{-1}(kT)^{-3}\propto\rho^{-2/3}T^{-3}$.

We have applied the data of spectral lines of some atomic elements (Nd~II, Fe~IV, and Fe~III) to test the derived formulae for the line expansion opacity and its Rosseland mean. The asymptotic behaviors of the Rosseland mean of the line expansion opacity described above are confirmed by the numerical results. The differences between the result given by our formulae and that given by the formulae of \cite{kar77} are outlined. Generally, the differences are evident for situations with $s\la 10$ (Fig.~\ref{mopc1}). We have also checked the variation of $\kappa_{\lex,R}$ with respect to the mass abundance $Y$ of the atomic element, which confirms the relation $\kappa_{\lex,R}\propto Y$ as $Y\rightarrow 0$ (Figs.~\ref{mopc1_fr} and \ref{mopc23_fr}). However, when $s$ is large and $Y$ is not small, $\kappa_{\lex,R}$ is not sensitive to the variation of $Y$ (Sec.~\ref{s_inf}).

To make the computation of the Rosseland mean of the line expansion opacity more efficient when the number of spectral lines is large, we have derived some approximation to the Rosseland mean, which works very well when $s$ is large. The approximation not only saves the time of computation considerably, but also leads to results with a high enough accuracy.

\acknowledgments

The author thanks an anonymous referee for a very enlightening and constructive report. This work was supported by the National Natural Science Foundation of China (Grant Nos. 11973014 and 11721303), and the National Basic Research Program (973 Program) of China (Grant No. 2014CB845800).

\end{document}